\documentclass[12pt]{article}
\usepackage{epsfig,rotating,lscape}
\usepackage{amssymb,latexsym,ifthen,bm,enumerate}
\usepackage{amsthm}
\usepackage{amsfonts}
\usepackage{natbib}
\usepackage{multirow}
\usepackage{epstopdf}
\usepackage{csquotes}
\usepackage{url} 
\usepackage{enumitem}
\usepackage{mathtools,leftidx}
\usepackage{xr}
\usepackage{soul}
\usepackage{tikz}
\usepackage{float}
\usepackage{orcidlink}
\usepackage{amsmath, amssymb}
\usepackage[ruled,vlined]{algorithm2e}
\usepackage{amsmath}
\usepackage{verbatim}
\usepackage{caption}
\usepackage{booktabs} 
\usetikzlibrary{shapes.geometric, arrows}

\hypersetup{
  colorlinks=true,
  linkcolor=black,
  urlcolor=black,
  citecolor=black,
  pdfborder={0 0 0} 
}
\makeatletter
\newcommand*{\addFileDependency}[1]{
  \typeout{(#1)}
  \@addtofilelist{#1}
  \IfFileExists{#1}{}{\typeout{No file #1.}}
}
\makeatother

\usepackage{array}
\newcolumntype{L}[1]{>{\raggedright\let\newline\\\arraybackslash\hspace{0pt}}m{#1}}
\newcolumntype{C}[1]{>{\centering\let\newline\\\arraybackslash\hspace{0pt}}m{#1}}
\newcolumntype{R}[1]{>{\raggedleft\let\newline\\\arraybackslash\hspace{0pt}}m{#1}}
\newcommand{\blind}{1}
\begin{document}

\if1\blind
{
  \bigskip
  \bigskip
  \bigskip
  \begin{center}
    {\LARGE \bf Deep Simulation-Based Inference for Inhomogeneous Bivariate Log-Gaussian Cox Processes}
    
    \vspace{1cm}
    {\large Qihan Zou$^{1}$, Yan Wang$^{1}$, Tingjin Chu$^{2}$, Zhendong Huang$^{1}$ \\}
    \vspace{0.5cm}
    {\it $^{1}$ School of Science, RMIT University\\  $^{2}$School of Mathematics and Statistics, The University of Melbourne}
  \end{center}
  \medskip
} \else {
  \bigskip
  \bigskip
  \bigskip
} \fi

\bigskip

\begin{abstract}
We propose a computationally efficient simulation-based estimation method with a two-step procedure for inhomogeneous bivariate Log-Gaussian Cox Processes. It combines classical Poisson estimation for the first-order parameters with simulation-based inference using neural networks for the latent field parameters. By separating the estimations, it reduces the complexity of high dimensional parameter estimation and the need for the simulation-based method to specify broad parameter ranges in the presence of covariates. In addition, we introduce two dimensional image inputs that enable the model to learn spatial information directly. Simulation results demonstrate that the proposed approach provides accurate estimates of the latent field parameters. We further illustrate the method's practical applicability using the gorilla dataset.
\end{abstract}

\noindent%
{\it Keywords: Simulation-based inference; Log-Gaussian Cox process; Likelihood-free inference; Neural networks.} 

\section{Introduction}
The Poisson point process is the fundamental model for modelling spatially random point patterns, characterised by complete spatial randomness and independent scattering \citep[see, e.g.][]{daley2003introduction,baddeley2016spatial}. The Poisson point process has been widely used in ecology, astronomy, seismology, epidemiology, insurance, criminology, environmental science, neuroscience, social networks analysis and other disciplines. While analytically tractable, its assumption of independence between points makes it unsuitable for phenomena exhibiting spatial aggregation or clustering. The log-Gaussian Cox process (LGCP) addresses this limitation by modelling the point process intensity as a realisation of an exponentiated Gaussian random field, therefore accommodating both first-order heterogeneity and second-order dependence within a flexible and computationally tractable framework \citep{moller1998log}. Despite its modelling flexibility, statistical inference for the LGCP, however, remains challenging. The likelihood involves integrating over an infinite-dimensional latent Gaussian field, leading to an analytically intractable likelihood function. 

A variety of methods have been developed that can be used to estimate LGCPs, including the composite likelihood \citep{guan2006composite}, Palm likelihood \citep{tanaka2008parameter}, Markov chain Monte Carlo (MCMC) \citep[see, e.g.][]{taylor2014inla,taylor2015bayesian,teng2017bayesian}, multilayer perceptron networks \citep{mateu2022spatial}, and Integrated Nested Laplace Approximation (INLA) \citep[see, e.g.][]{rue2009approximate,taylor2014inla,flagg2023integrated}. These methods have been commonly used in fitting and making inferences for LGCPs. However, these classical inference strategies are either theoretically difficult for practitioners to implement correctly or computationally prohibitive in realistic applications. Minimum-contrast estimation \citep[see, e.g.][]{diggle1984monte, waagepetersen2007estimating} based on summary statistics, such as the pair-correlation function or the K-function, provides a more tractable alternative, but its accuracy is often limited, and it does not scale well to higher-dimensional parameter spaces. Recent work by \cite{dovers2024fitting} proposed a method that uses well-established generalised additive model software to fit log-Gaussian Cox processes, making the estimation process more accessible. 

\cite{diggle1983bivariate} develop bivariate Cox process models for analysing dependence between two spatial point patterns using second-order summary functions, such as cross-type K-functions. However, their main development is limited to stationary settings. \cite{moller1998log} develop a formal framework mainly for univariate and multivariate stationary log-Gaussian Cox process models for clustered spatial point patterns. They note that likelihood-based inference is analytically intractable and that pseudo-likelihood is not useful, and therefore estimate LGCP parameters using summary statistics, such as the empirical pair correlation function, through a minimum contrast method. \cite{brix2001spatiotemporal} extend log-Gaussian Cox processes to spatio-temporal point process data by modelling the latent intensity surface as a space-time Gaussian process, and estimates the spatial covariance parameters using a minimum contrast method. However, their method is restricted to univariate LGCPs. A more general multi-type log-Gaussian Cox process set-up is present in \cite{waagepetersen2016analysis} and \cite{choiruddin2020regularized}. \cite{waagepetersen2016analysis} proposed a least-squares estimation method for multivariate LGCPs based on pair and cross-pair correlation functions. \cite{choiruddin2020regularized} extended this approach by developing a regularised least-squares method to improve stability in highly multivariate LGCPs. However, these approaches require cross-validation or regularisation procedures to determine the latent factor structure, introducing an additional model-selection step that is unnecessary in our bivariate LGCP setting. Recently, \cite{zhu2025minimum} proposed a minimum contrast method based on a Q-function matrix for homogeneous multivariate point processes, supported by rigorous mathematical theory. However, their method does not address the inhomogeneous case.

Simulation-based inference (SBI) enables parameter estimation in models with intractable likelihoods by learning from simulated data rather than closed-form likelihood expressions \citep{cranmer2020frontier, lueckmann2021benchmarking}. This makes SBI particularly suitable for complex spatial point process models such as the LGCP.  Recently, \cite{vihrs2022using} introduced a neural-network simulation-based approach for parameter estimation in homogeneous LGCPs and several other spatial point processes. We name this class of approaches Deep Simulation-Based Inference (DSBI). The approach demonstrated how deep learning can successfully recover parameters from simulated summary statistics without relying on closed-form likelihoods. While recent work has demonstrated the potential of SBI for parameter estimation in homogeneous LGCP, many real-world spatial point patterns exhibit substantial spatial inhomogeneity driven by environmental or contextual covariates. The inhomogeneous LGCP includes both a deterministic first-order intensity component and a stochastic latent field describing residual clustering. Extending SBI to inhomogeneous LGCPs is therefore of practical importance but at the same time introduces additional challenges. Firstly, the inhomogeneous LGCP introduces potential confounding between covariate-driven intensity variation and dependence induced by the latent Gaussian field. Furthermore specifying an appropriate sampling range for the covariate parameters when generating training simulations is demanding. If these ranges are too wide, the simulated patterns may contain unrealistically many points or none at all, because the mean trend heavily drives the expected events' intensity. Additionally, including more covariates in the LGCP mean function substantially enlarges the neural-network parameter space, making training unstable and computationally demanding. On the other hand, extending the \texttt{DSBI} method from univariate to bivariate cases requires a richer class of summary statistics, as the number of latent field parameters increases substantially and identifiability becomes weaker. Therefore, introducing a new class of summary statistics is urgent and crucial.

The two-step estimation procedure for inhomogeneous Neyman–Scott processes was introduced by \cite{waagepetersen2007estimating}. In the first step, the (first-order) regression parameters are estimated using a Poisson-based estimating function. In the second step, the second-order parameters are estimated by minimum contrast estimation. This idea was later formalised by \cite{waagepetersen2009two}. They developed a general two-step estimation framework for inhomogeneous spatial point processes with rigorous theoretical justification. They show that the Poisson estimating function for the regression parameters is unbiased and establish asymptotic normality of the resulting two-step estimator. Building on the DSBI \citep{vihrs2022using}, we introduce a new DSBI specifically designed for inhomogeneous bivariate LGCPs. Specifically, our approach applies a two-step estimation that separates the estimation of the first-order and second-order components of the model: the first-order intensity associated with spatial covariates is estimated using a classical statistical estimator, after which a simulation-based inference with a neural network is employed to estimate the parameters of the latent Gaussian field conditional on the estimated intensity. This hybrid strategy reduces the large number of parameters problem faced by \citet{vihrs2022using}, improves estimator stability, and allows the neural network to focus on learning the complex second-order dependence structure. Once trained, the neural network provides instant, likelihood-free estimation for new data \citep{cranmer2020frontier,zammit2025neural}. In addition, we introduce a new class of input, which we call spatial structure input, and use them as 2D image inputs for the \texttt{DSBI} method. We use the 2D count image as an example of this new class of input to demonstrate its importance in the bivariate setting. Moreover, we provide a general workflow for practical applications, such as how to select appropriate parameter ranges for the scale parameter and validate the trained network on real spatial datasets. 

The structure of the remaining sections is summarised as follows. In Section 2, we propose a neural network-based method. Section 3 presents the simulation setup and the results of the simulation studies. In Section 4, we apply our method to the gorilla data set and evaluate the performance of various approaches. Finally, Section 5 provides discussions and conclusions.

\section{Method}
\subsection{Bivariate log-Gaussian Cox Process}
The log-Gaussian Cox process (LGCP) \citep[see, e.g.,][]{moller1998log,baddeley2016spatial,moraga2023spatial} is a Poisson process with a non-negative random intensity function $\Lambda(\bm{s})$, where the log-intensity is modelled as a Gaussian random field. A spatial point pattern $X$ containing points of different types is called a multi-type point pattern. 

Consider a bivariate log-Gaussian Cox process \citep[see, E.g.][]{diggle1983bivariate} for types $p = 1,2$, we have two types of point patterns and $X=\{X_1, X_2\}$ and the following intensity structure, $$\log\{\Lambda_p(\bm s)\} = \bm Z(\bm s)\bm \beta_p + Y(\bm s) + U_p(\bm s),$$
where $Y(\bm s)$ is the shared field that affects all types of processes.

Specifically, $Y(\bm s)$ is a zero-mean latent Gaussian random field with stationary and isotropic covariance, $\text{cov}\{Y(\bm s), Y(\bm s + \bm h)\} = c_Y(\bm h;\bm \theta_Y)$, where $\bm h \in \mathbb{R}^2$ and $\bm \theta_Y$ is the second-order parameter of interest. Throughout the paper, we assume the covariance $c_Y(\bm h;\bm \theta_Y) = \sigma_Y^2 \exp(-\|\bm h\|/\xi_Y)$, where $\bm \theta_Y = (\sigma_Y,\xi_Y)$ and $\|\bm h\|$ denotes the Euclidean norm of $\bm h$. Similarly, $U_p(\bm s)$ is the stationary and isotropic latent Gaussian random field, which only affects individually the type $p$ process, with zero-mean and covariance $c_{U_p}(\bm h; \bm \theta_{U_p}) = \sigma_{U_p}^2 \exp(-\|\bm h\| /\xi_{U_p})$ and $\bm \theta_{U_p} = (\sigma_{U_p},\xi_{U_p})$. Finally, $\bm Z(\bm s)$ represents deterministic covariates and $\bm \beta_p$ are the associated coefficients.

For a multivariate LGCP, the marginal mean intensity is $$ E{\{\Lambda_p(\bm s)\}}=\exp\left(\bm Z(\bm s) \bm\beta_p+\frac{1}{2}Var\{Y(\bm s)+U_p(\bm s)\}\right). $$
If we use the centred LGCP form
$$\log\{\Lambda_p(\bm s)\} = \bm Z(\bm s) \bm\beta_p + Y(\bm s) + U_p(\bm s)-\frac{1}{2}\sigma_Y^2-\frac{1}{2}\sigma^2_{U_p}$$
then $$ E{\{\Lambda_p(\bm s)\}}=\exp\{\bm Z(\bm s) \bm\beta_p\}. $$ Using the centred LGCP formulation, the resulting fitted intensities will provide estimates of the mean trend functions.

\subsection{Inputs of Neural Network}
We consider three classes of inputs to our proposed method: first-order estimates inputs $S_1$, second-order summary statistics inputs $S_2$, and the proposed spatial structure inputs $S_3$. We denote this group of three input classes as $S = (S_1, S_2, S_3)$.

First-order estimates inputs $S_1$ that describe global properties and are most closely related to the mean trend of the point patterns. Empirical second-order summary statistics often rely on estimates of mean trend parameters. Consequently, although they are not directly informative about second-order parameters, they remain important for their estimation. We include estimates of mean trend parameters $\hat{\bm{\beta}}_p$ for each type $p=1,2$ as inputs. They serve as complementary inputs to the other inputs, providing high-level global summarised information about the observed point patterns. Therefore, we have first-order estimates inputs $S_1 = {\hat{\bm{\beta}}_p}$ where $p=1,2$.

The second-order summary statistics inputs $S_2$ play a central role in \texttt{DSBI} as it direct relate to the latent field parameters. The inhomogeneous K-function is the expected total weight of all locations $\bm{s}_i$ within a radius $r$ of a location $\bm{s}_j$, where $i \neq j$, and each location $\bm{s}_i$ is assigned a weight equal to the reciprocal of its intensity, $1/\Lambda(\bm{s}_i)$. For the simulation-based inference, we are interested in the empirical summary statistics generated from the simulation processes. The empirical estimator of the inhomogeneous K-function \citep[see, e.g.,][]{baddeley2000non, baddeley2016spatial} is defined as $$\widehat{K}_{\text{inhom}}(r) = \frac{1}{|W|} \sum_i \sum_{j \neq i}  \frac{\mathbf{1}\{ \|\bm h_{ij}\| \le r \}  e(\bm{s}_i, \bm{s}_j; r)}{\hat \lambda(\bm{s}_i)\hat \lambda(\bm{s}_j)},$$ where the points $\bm s$ are only observed in a bounded window $W \subset \mathbb{R}^2$, $|W|$ is Lebesgue measure (area) of $W$, $\hat \lambda(.)$ is the estimated first-order intensity, $r$ is the chosen radius, $\bm h_{ij} = \bm{s}_i - \bm{s}_j$ is the distance between the points $\bm{s}_i$ and $\bm{s}_j$, and $e(\bm{s}_i, \bm{s}_j; r)$ is the edge correction weight. Following \citet{baddeley2000non}, the inhomogeneous empirical L-function $$\widehat L_{\text{inhom}}(r) = \sqrt{\frac{\widehat K_{\text{inhom}}(r)}{\pi}} $$ is a transformation of the inhomogeneous K-function. Under an inhomogeneous Poisson process with intensity $\lambda(\bm s)$, the expected L-function $L_{\text{inhom}}(r) -r = 0$. Departures from zero indicate residual spatial dependence beyond that explained by the intensity. In particular, $L_{\text{inhom}}(r) -r > 0$ suggests clustering and $L_{\text{inhom}}(r) -r < 0$ suggests repulsion or regularity. We use the following second-order summary statistics in this work, $S_2 = \{\widehat L_{11}(r) - \widehat L_{12}(r), \: \widehat L_{22}(r) - \widehat L_{12}(r), \: \widehat L_{12}(r) - r \}$ where $\widehat L_{11}$ and $\widehat L_{22}$ denote the estimated within-type L-functions for type 1 and type 2. $\widehat L_{12}$ denotes the estimated cross-types L-function for type 1 and type 2. 

To better retain spatial information in bivariate and multivariate point patterns, we introduce the spatial structure inputs $S_3$, based on standardised count images. Traditional second-order summary statistics are highly aggregated. As a result, they lose important information about the spatial arrangement of points. Image-based inputs help reduce this loss by preserving the spatial layout of the observed pattern and allowing the model to learn more complex spatial dependencies. 

Let the observation window $W$ be partitioned into non-overlapping grid cells $g_1,\ldots,g_M$. Let $N_{11}(g_m)$ and $N_{22}(g_m)$ denote the number of points observed in cell $g_m$ associated with the types $1$ and $2$, respectively. Let $ N_{12}(g_m)=N_{11}(g_m)+N_{22}(g_m)$ be the total observed (pooled) points in $g_m$. Let $\hat \lambda_{11}(s)$, $\hat \lambda_{22}(s)$ and $\hat \lambda_{12}(s) = \hat \lambda_{11}(s) + \hat \lambda_{22}(s)$ be the corresponding estimated mean intensity. We define three standardised count images as follows. For $1\leq p \leq q \leq 2$, let 
$$I_{pq}(g_m) = \frac{N_{pq}(g_m)-E_{pq}(g_m)}{\sqrt{E_{pq}(g_m)}}, \ \ \text{where} \: \: E_{pq}(g_m) = \int_{g_m} \hat \lambda_{pq}(s)\,\mathrm{d}s. 
$$ 
Define the spatial structure input $S_3=\{I_{11}, I_{22}, I_{12}\}$, where $I_{pq}$ is an (vectorized) image, collecting $I_{pq}(g_m)$ for $m=1,\dots,M$.

\subsection{Deep Simulation-Based Inference Algorithm}
\label{subsec: IDSBIA}
For the bivariate LGCP, we separate the estimation of the first-order trend parameters $\bm \beta_p$ from that of the latent field parameters following the two-step estimation procedure introduced by \cite{waagepetersen2007estimating} and \cite{waagepetersen2009two}. 

In the first stage, we estimate the first-order trend parameters for each type using separate inhomogeneous Poisson point process models, implemented through \texttt{ppm()} in the \texttt{spatstat} package. Under the centred LGCP formulation, the resulting fitted intensities provide estimates of the mean trend functions of the two types. These estimated intensity surfaces are subsequently treated as known and are used in the construction of empirical inhomogeneous second-order summary statistics $S_2$, such as the within-type and cross-type L-functions, which capture residual spatial dependence after accounting for large-scale intensity variation. The estimated trend parameters $\hat\beta_p$, empirical second-order summary statistics, along with the standardised count image provide complementary information on the first-order and second-order structure of the point patterns. These quantities are therefore used jointly as inputs to the second-stage neural network for estimating the covariance parameters of the bivariate LGCP. The Deep simulation-based inference algorithm is shown in Algorithm~\ref{alg:alg1}.

\begin{figure}[H]
\centering
\begin{tikzpicture}[
    font=\small,
    >=latex,
    block/.style={
        draw,
        rounded corners=2pt,
        align=center,
        minimum width=3.1cm,
        minimum height=0.82cm
    },
    arrow/.style={->, line width=0.7pt}
]
\draw[rounded corners=4pt, line width=0.8pt] (-6.1,3.9) rectangle (4.3,-8.4);
\draw[dashed, rounded corners=4pt, line width=0.8pt] (5.4,2.8) rectangle (10.8,-5.5);
\node[font=\bfseries] at (-0.9,3.45) {Neural Network Design};
\node[font=\bfseries] at (8.1,2.35) {DSBI Workflow};
\node[font=\bfseries] at (-3.5,2.75) {$S_2$ Branch};
\node[font=\bfseries] at (1.5,2.75) {$S_3$ Branch};
\node[font=\bfseries] at (3,-4) {$S_1$ Branch};
\node (pcf0) at (-3.5,2.15) {Input: $S_2$};
\node[block] (pcf1) at (-3.5,1) {1D Convolution layer\\+ 1D max pooling};
\node[block] (pcf2) at (-3.5,-0.45) {1D Convolution layer\\+ 1D max pooling};
\node[block] (pcf3) at (-3.5,-1.75) {1D Convolution layer};
\node (img0) at (1.5,2.15) {Input: $S_3$};
\node[block] (img1) at (1.5,1) {2D Convolution layer\\+ 2D max pooling};
\node[block] (img2) at (1.5,-0.45) {2D Convolution layer\\+ 2D max pooling};
\node[block] (img3) at (1.5,-1.75) {2D Convolution layer};
\node[block] (img4) at (1.5,-3) {2D average pooling};
\node (scalar0) at (3,-4.5) {Input: $S_1$};
\node[block, minimum width=5.4cm] (cat) at (-1.8,-4.5) {concatenation};
\node[block, minimum width=5.4cm] (d1) at (-1.8,-5.6) {Hidden layer (128 neurons), ReLU};
\node[block, minimum width=5.4cm] (d2) at (-1.8,-6.7) {Hidden layer (64 neurons), ReLU};
\node[block, minimum width=5.4cm] (out) at (-1.8,-7.8) {Output: $\bm\theta$};
\draw[arrow] (pcf0) -- (pcf1);
\draw[arrow] (pcf1) -- (pcf2);
\draw[arrow] (pcf2) -- (pcf3);
\draw[arrow] (pcf3) -- (cat);
\draw[arrow] (img0) -- (img1);
\draw[arrow] (img1) -- (img2);
\draw[arrow] (img2) -- (img3);
\draw[arrow] (img3) -- (img4);
\draw[arrow] (img4) -- (cat);
\draw[arrow] (scalar0) -- (cat);
\draw[arrow] (cat) -- (d1);
\draw[arrow] (d1) -- (d2);
\draw[arrow] (d2) -- (out);
\node[block, minimum width=3.8cm] (xobs) at (8.1,1.45) {Input: $X^{(obs)}$};
\node[block, minimum width=3.8cm] (first) at (8.1,0) {First-order estimation};
\node[block, minimum width=4.5cm] (stats) at (8.1,-1.45) {$S(X^{(obs)}) = (S_1, S_2, S_3)$};
\node[block, minimum width=4.5cm] (pre) at (8.1,-2.9) {Pre-trained DSBI model};
\node[block, minimum width=3.8cm] (thetahat) at (8.1,-4.35) {Output: $\hat{\bm\theta}$};
\draw[arrow] (xobs) -- (first);
\draw[arrow] (first) -- (stats);
\draw[arrow] (stats) -- (pre);
\draw[arrow] (pre) -- (thetahat);
\draw[arrow, dashed, line width=0.9pt] (4.3,-2.9) -- (pre.west);
\end{tikzpicture}
\caption{The \texttt{DSBI} workflow and the neural network design.}
\label{fig:method1}
\end{figure}
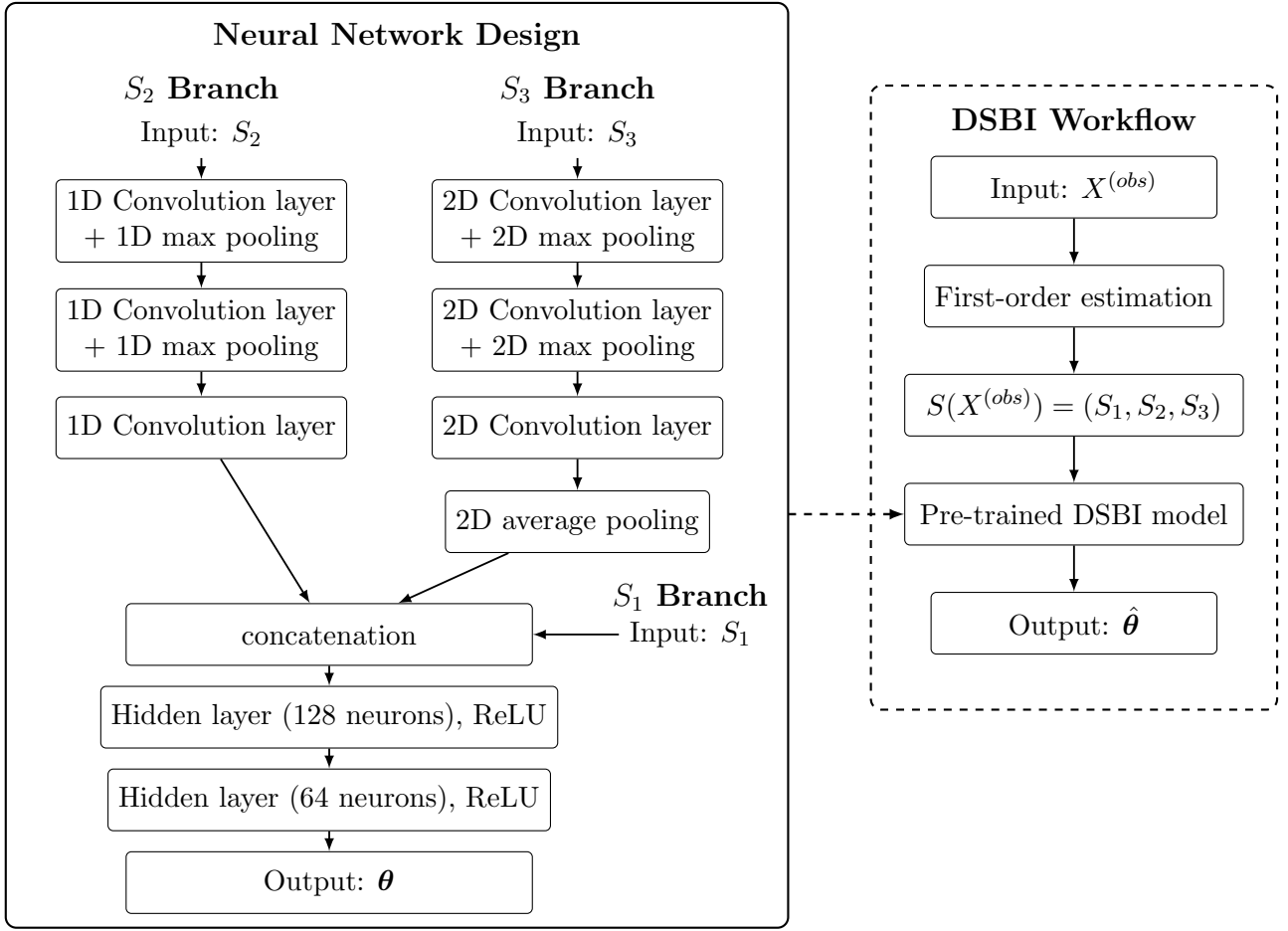

The overall \texttt{DSBI} workflow and the details of neural network design are shown in Figure~\ref{fig:method1}. For the $S_2$ branch and the layers after concatenation, we follow the model architecture designed in \cite{vihrs2022using}. For the spatial structure inputs ($S_3$ branch), we use 2-dimensional Convolution layers as $S_3$ is an image input. Max pooling and average pooling are used to reduce the size of vectors or images while capturing important information from the inputs. Max pooling identifies the strongest signals, whereas adaptive average pooling computes the average to represent the overall structure. These methods help extract meaningful features from the inputs. After extracting important information from $S_2$ and $S_3$, these features are concatenated with the first-order summary statistics $S_1$ and passed to a multilayer Perceptron with two hidden layers with Rectified Linear Unit (ReLU) activation function \citep{agarap2018deep} to model the complex relationship with the response variable. To ensure the positivity of the latent field parameters, we apply a square root transformation to $\bm\theta$ and square the model outputs.

\begin{algorithm}[H]
\caption{Deep simulation-based inference}
\textbf{Training stage}
\begin{enumerate}
\item For each simulation $b=1,\dots,B$, for some user-specified constant $B$, sample $ \bm \beta^{(b)} = (\bm \beta^{(b)}_1, \bm \beta^{(b)}_2)$ and $\bm \theta^{(b)} = (\bm \theta_Y^{(b)}, \bm \theta_{U_1}^{(b)}, \bm \theta_{U_2}^{(b)})$ from pre-defined uniform distributions.
\item For each simulation $b$:
\begin{enumerate}
\item Sample point pattern $X^{(b)}=\{X^{(b)}_1, X^{(b)}_2\}$ from bivariate LGCP$(\bm \beta^{(b)}, \bm \theta^{(b)})$.
\item Given $X^{(b)}$, fit an inhomogeneous Poisson process model to obtain $\hat{\bm \beta}^{(b)}_{p}$ for each type.
\item Obtain inputs $S(X^{(b)}) = \{S^{(b)}_1, S^{(b)}_2, S^{(b)}_3\}$:
\begin{enumerate}
    \item Obtain first-order estimates inputs $S^{(b)}_1 = (\hat{\bm \beta}^{(b)}_1, \hat{\bm \beta}^{(b)}_2)$.
    \item Compute the second-order summary statistics inputs $S^{(b)}_2 = \{\widehat L^{(b)}_{11}(r) - \widehat L^{(b)}_{12}(r), \: \widehat L^{(b)}_{22}(r) - \widehat L^{(b)}_{12}(r), \: \widehat L^{(b)}_{12}(r) - r \}$ by plugging in $\hat{\bm \beta}^{(b)}_p$.
    \item Obtain spatial structure inputs $S^{(b)}_3 = \{I^{(b)}_{11}, I^{(b)}_{22}, I^{(b)}_{12}\}$.
\end{enumerate}
\end{enumerate}
\item Train a neural network using inputs $S(X^{(b)})$ and targets $\bm \theta^{(b)}$, for all $b= 1,\dots, B$.
\end{enumerate}
\textbf{Estimation stage}
\begin{enumerate}
\item Given an observed data $X^{(\text{obs})}$, fit an inhomogeneous Poisson process model to obtain $\hat{\bm \beta}^{(\text{obs})}_{p}$ for each type.
\item Compute inputs $S(X^{(\text{obs})})$ as 2 (c) in Training stage.
\item Input $S(X^{(\text{obs})})$ into the trained network to obtain estimates of $\bm \theta$.
\end{enumerate}
\label{alg:alg1}
\end{algorithm}

\section{Simulation Study}
We conducted two simulation studies to evaluate the performance of the proposed \texttt{DSBI} method. In Simulation Study 1, we considered a homogeneous bivariate LGCP, and compared \texttt{DSBI} with the homogeneous minimum contrast method (\texttt{MC}) \citep{zhu2025minimum} and the Integrated Nested Laplace Approximation (\texttt{INLA}) \citep[see, e.g.,][]{rue2009approximate,lindgren2015bayesian} as benchmarks. In Simulation Study 2, we considered an inhomogeneous bivariate LGCP and evaluated the performance of \texttt{DSBI} method.

\subsection{Homogeneous Bivariate LGCP}
For types $p=1,2$, we considered a homogeneous bivariate LGCP with intensity, $$\Lambda_p(\bm s) = \exp \{\mu_{p} + Y(\bm s) + U_p(\bm s) - \frac{1}{2}\sigma_Y^2 - \frac{1}{2}\sigma_{U_p}^2\},$$ where the shared field $Y(\bm s)$ was a zero-mean latent Gaussian random field with exponential covariance $\sigma^2_Y \exp(-\|\bm h\|/\xi_Y)$. Each of the individual fields $U_p(\bm s)$ was a zero-mean latent Gaussian random field with exponential covariance $\sigma^2_{U_p} \exp(-\|\bm h\|/\xi_{U_p})$. 

The simulated window was set to $[0,1] \times [0,1]$ and the window was discretised into a $50 \times 50$ grid. To construct the training data, we sampled the parameters uniformly from the training ranges shown in Table~\ref{tab:hom_param1} and generated 100,000 realisations for the training set. We used a batch size of 100, 30 epochs and a learning rate of 1e-3. To ensure positivity of the variance and scale parameters, we applied a square-root transformation to the response variables during training and then squared the predicted values to obtain the final estimates. 

The true parameter values in the testing set were fixed, as shown in Table~\ref{tab:hom_param1}. We conducted 500 simulations to evaluate the performance of the proposed method against the minimum contrast method of \citet{zhu2025minimum}, implemented following their procedure, and \texttt{INLA} \citep[see, e.g.,][]{rue2009approximate,lindgren2015bayesian}. We used bias, defined as the difference between the average estimate and the true value, root mean squared error (RMSE) and mean absolute percentage error (MAPE) to evaluate the performance of the methods. The MAPE was defined as $\text{MAPE} =  \frac{1}{n} \sum^n_{i=1} |\frac{\theta_i - \hat{\theta_i}}{\theta_i}|$, where $\theta_i$ was the actual value and $\hat{\theta_i}$ was the predicted value.

The \texttt{DSBI} model was trained over a predefined parameter space, and therefore its predictions are primarily intended for parameter values within these training ranges. To ensure a fair comparison across different methods, we restricted the parameter space considered by the classical approaches to the same ranges whenever possible. Specifically, for the minimum contrast (\texttt{MC}) method, we used the \texttt{L-BFGS-B} algorithm in \texttt{optim()} with lower and upper bounds matching the \texttt{DSBI} training ranges. For \texttt{INLA}, explicit parameter bounds cannot be imposed directly, so we specified priors based on the empirical distributions of the simulated training parameters. As \texttt{INLA} may still produce estimates outside the training domain, these estimates were excluded from the comparison, as they lie outside the parameter space on which the \texttt{DSBI} model was trained.

\begin{table}[H]
\centering
\resizebox{\textwidth}{!}{
\begin{tabular}{lcccccccc}
\hline
 & $\mu_{1}$ & $\mu_{2}$ & $\sigma_Y$ & $\xi_Y$ 
 & $\sigma_{U_1}$ & $\sigma_{U_2}$ & $\xi_{U_1}$ & $\xi_{U_2}$ \\
\hline
Testing value     
& 6 & 6 & 1.5 & 0.2 & 1 & 1 & 0.15 & 0.15 \\
Training range 
& $[5,7]$ & $[5,7]$ & $[0.5,2]$ & $[0.001,0.3]$ 
& $[0.5,2]$ & $[0.5,2]$ & $[0.001,0.3]$ & $[0.001,0.3]$ \\
\hline
\end{tabular}
}
\caption{True parameter values and training ranges for homogeneous bivariate LGCP.}
\label{tab:hom_param1}
\end{table}

\begin{table}[H]
\centering
\begin{tabular}{cc|ccc|ccc|ccc}
\hline
 & & \multicolumn{3}{c|}{DSBI} & \multicolumn{3}{c|}{MC} & \multicolumn{3}{c}{INLA}\\
& True   & Bias & RMSE & MAPE  & Bias & RMSE& MAPE & Bias & RMSE & MAPE \\
\hline
$\sigma_Y$ & 1.5 & -0.082 &0.211& 10.713& -0.303 &0.476 &26.357& -0.069& 0.274& 14.410\\
$\sigma_{U_1}$& 1 & 0.048 &0.216& 17.170& -0.163 &0.336 &28.562& -0.029& 0.253& 20.629 \\
$\sigma_{U_2}$& 1 & 0.030 &0.208& 16.803& -0.163 &0.328 &27.848& -0.027& 0.259& 21.108\\
\hline
$\xi_Y$ & 0.2& -0.007& 0.031& 11.965 & -0.046 &0.109 &47.330& 0.054& 0.064& 28.562\\
$\xi_{U_1}$& 0.15 & 0.014& 0.035& 18.813& -0.045 &0.103 &61.701& 0.070& 0.086 &50.529\\
$\xi_{U_2}$& 0.15 & 0.016& 0.035& 19.272& -0.052 &0.101 &60.137& 0.067& 0.083 &48.504\\
\hline
\end{tabular}
\caption{Comparison of parameter estimation performance across methods. The table reported bias along with the root mean squared error (RMSE) and mean absolute percentage error (MAPE) as a percentage (\%).}
\label{tab:hom_comparison}
\end{table}

\begin{figure}[H]
    \centering
    \includegraphics[width=1\linewidth]{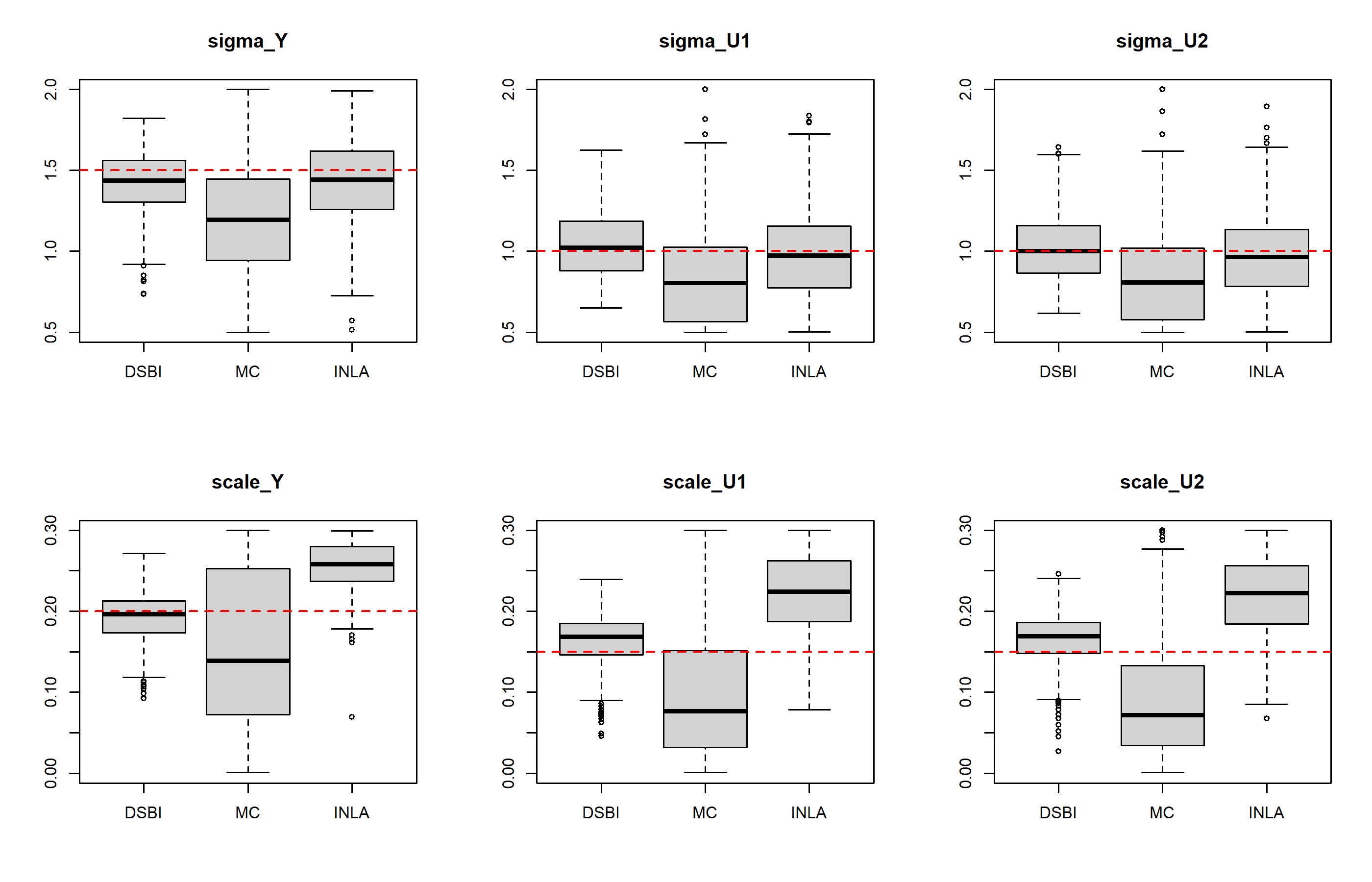}
    \caption{Boxplots comparing the estimates of the second-order parameters obtained using \texttt{DSBI}, \texttt{MC} and \texttt{INLA}.}
    \label{fig:hom_boxplot}
\end{figure}

Based on Table~\ref{tab:hom_comparison} and Figure~\ref{fig:hom_boxplot}, \texttt{DSBI} provided the most accurate estimates overall. For the variance parameters, \texttt{DSBI} had the smallest RMSEs for $\sigma_Y$, $\sigma_{U_1}$ and $\sigma_{U_2}$, and its biases were close to zero. The estimate of $\xi_Y$ had a small bias, while the estimates of $\xi_{U_1}$ and $\xi_{U_2}$ were slightly overestimated. In addition, the computational time for each method is reported in Appendix~\ref{sec:appendix_time}.

Although the RMSEs of the scale parameters were smaller than those of the variance parameters, this mainly reflected the difference in their numerical scales. The MAPE, a scale-independent measure of estimation accuracy, showed consistently larger relative errors for the scale parameters across all three methods. These results suggested that the scale parameters were intrinsically more difficult to estimate than the variance parameters. However, the discrepancy between the estimation accuracies of the scale and variance parameters was substantially smaller for \texttt{DSBI} than for the other two competing methods. This demonstrated that \texttt{DSBI} maintained more uniform performance across parameters with different characteristics.

The \texttt{MC} method was a relatively lightweight and fast method that relied only on the single observed data set and it tended to underestimate most second-order parameters, especially the scale parameters. This was also reflected in its larger RMSE and MAPE values. In particular, the estimates of $\xi_{U_1}$ and $\xi_{U_2}$ from \texttt{MC} showed substantial downward bias and large relative errors. The \texttt{INLA} method performed reasonably well for the variance parameters, but it tended to overestimate the scale parameters. This could be seen clearly from the boxplots, where the \texttt{INLA} estimates for $\xi_Y$, $\xi_{U_1}$, and $\xi_{U_2}$ were mostly above the true values. These results suggested that, by using a large amount of training data, \texttt{DSBI} gave more stable and accurate estimates with relatively small variability compared with both benchmark methods in this homogeneous bivariate LGCP setting. In addition, results across methods for the larger window $[0,1.5]^2$ are reported in Appendix~\ref{sec: sim1_larger_W}. 

\subsection{Inhomogeneous Bivariate LGCP}
\label{sec:Inhomogeneous Bivariate LGCP}
\begin{figure}[H]
    \centering
    \includegraphics[width=0.45\linewidth]{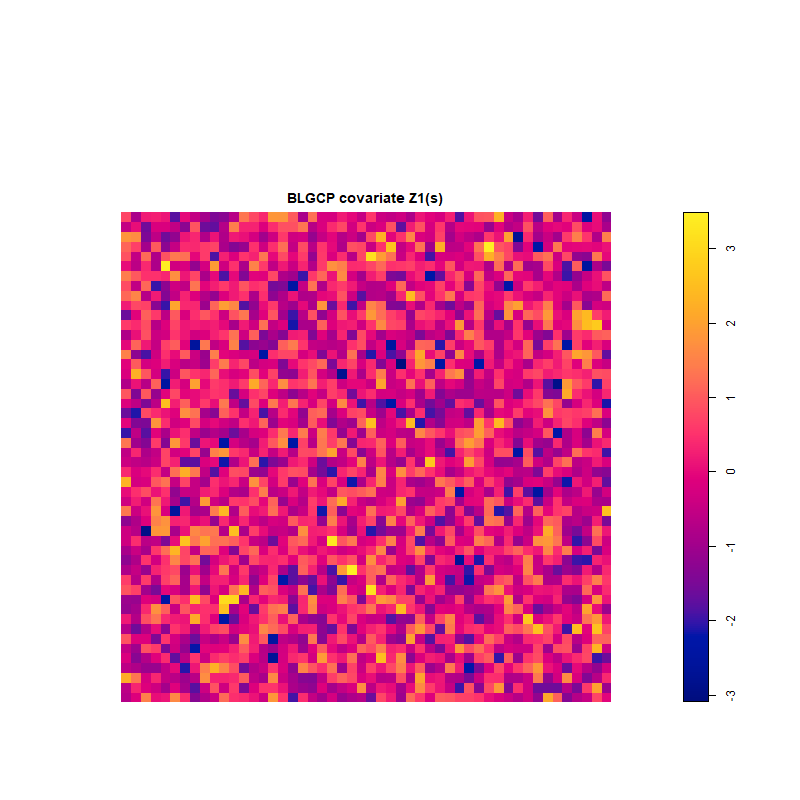}
    \quad
    \includegraphics[width=0.45\linewidth]{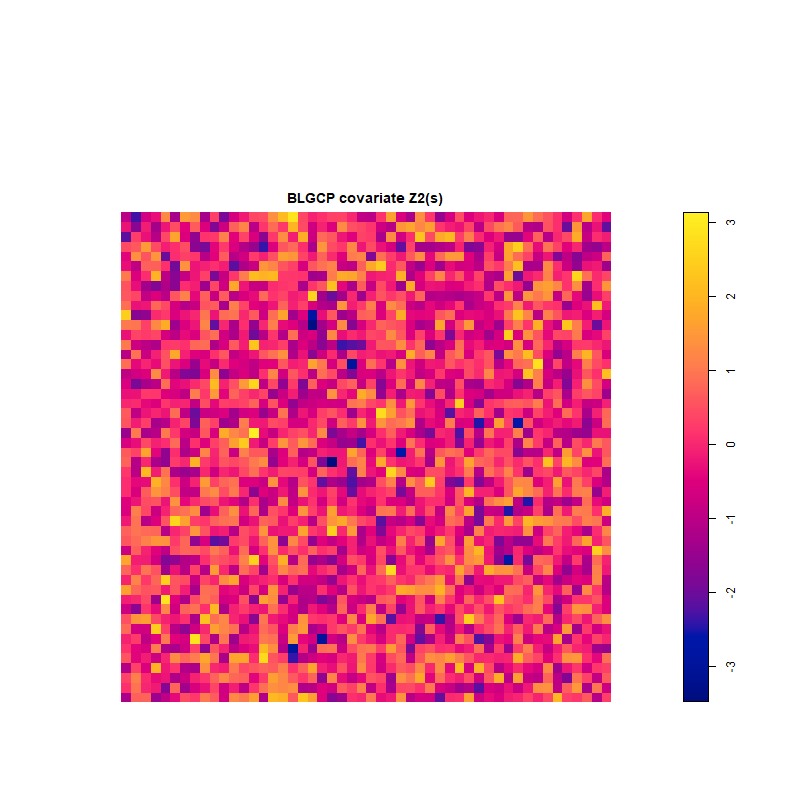}
    \caption{The fixed and known covariates $Z_1(s)$ and $Z_2(s)$ in the BLGCP model.}
    \label{fig:inhom_covariate_fields}
\end{figure}

We considered a bivariate inhomogeneous LGCP with the following intensity function for $p=1,2$, $$\Lambda_p(s) = \exp \{\beta_{0,p} + \beta_{1,p} Z_1(s) + \beta_{2,p} Z_2(s) + Y(s) + U_p(s) - \frac{1}{2}\sigma_Y^2 - \frac{1}{2}\sigma_{U_p}^2\},$$ where the specifications of the shared field $Y(s)$ and the individual fields $U_p(s)$ were the same as those in Simulation study for homogeneous bivariate LGCP. In addition, $\beta_{0,p}$, $\beta_{1,p}$ and $\beta_{2,p}$ were first-order parameters of the intercept and fixed and known covariates $Z_1(s)$ and $Z_2(s)$ from two independent realisations of a mean-zero Gaussian random field (see Figure~\ref{fig:inhom_covariate_fields}). We considered two simulation windows $W_1$ and $W_2$, which were set to $[0,1] \times [0,1]$ and $[0,1.5] \times [0,1.5]$, respectively. Each window was discretised into a $50 \times 50$ grid.

For constructing the training and testing data, we sampled the parameters $\beta_{0,p}$, $\beta_{1,p}$ and $\beta_{2,p}$, as well as the latent field parameters $\sigma_Y$, $\xi_Y$, $\sigma_{U_p}$, and $\xi_{U_p}$, within the intervals shown in Table~\ref{tab:inhom_param}. The $100000$ realisations for the training set and $5000$ realisations for the testing set were sampled independently. We used a batch size of 100, 30 epochs and a learning rate of 1e-3. Similar to simulation study for homogeneous bivariate LGCP, we applied a square-root transformation to ensure positivity of the variance and scale parameters. 

\begin{table}[H]
    \centering
    \begin{tabular}{cccccc}
    \hline
    Sampling Interval& $\beta_{0,p}$ & $\beta_{1,p}$ & $\beta_{2,p}$ & $\sigma_Y$ \& $\sigma_{U_p}$ & $\xi_Y$ \& $\xi_{U_p}$\\
    \hline
    Training &  $ [5,7]$ &  $ [-1,1]$ & $ [-1,1]$& $[0.5, 2]$ & $[0.001, 0.2]$\\
    Testing  &  $ [5.5, 6.5]$ & $ [-0.5,0.5]$ & $ [-0.5,0.5]$& $[1, 1.5]$ & $[0.07, 0.13]$\\
    \hline
    \end{tabular}
    \caption{Sampling intervals for the parameters used in the simulation study. The table reports the training and testing ranges for the regression coefficients, the shared latent field parameters $(\sigma_Y, \xi_Y)$, and the individual latent field parameters $(\sigma_{U_p}, \xi_{U_p})$, for $p=1,2$.} 
    \label{tab:inhom_param}
\end{table}

\begin{table}[H]
\centering
\begin{tabular}{c|ccc|ccc}
\hline
& \multicolumn{3}{c|}{DSBI, $W_1 = [0, 1]^2$} & \multicolumn{3}{c}{DSBI, $W_2 = [0, 1.5]^2$}  \\
Parameter & Bias & RMSE  & MAPE  & Bias & RMSE  & MAPE  \\
\hline
$\sigma_Y$ & 0.078 & 0.184 &12.309& 0.028 & 0.121 & 7.893\\
$\sigma_{U_1}$ & -0.088& 0.219 &14.338&-0.044& 0.143 & 9.154\\
$\sigma_{U_2}$ &-0.064 & 0.199 &13.120&-0.007& 0.140 & 9.105\\
\hline
$\xi_Y$ & 0.009 & 0.023 &19.789& 0.005& 0.016 &14.469\\
$\xi_{U_1}$ & 0.006&0.025 &20.801&-0.002& 0.017 &13.978\\
$\xi_{U_2}$ & 0.009&0.028 &23.494& 0.002& 0.017 &14.687\\
\hline
\end{tabular}
\caption{Comparison of parameter estimation performance for the proposed method with different windows. The table reports the bias (Bias), root mean squared error (RMSE) and mean absolute percentage error (MAPE) as a percentage (\%).}
\label{tab:comparison_table}
\end{table}

\begin{figure}[H]
    \centering
    \includegraphics[width=0.7\linewidth]{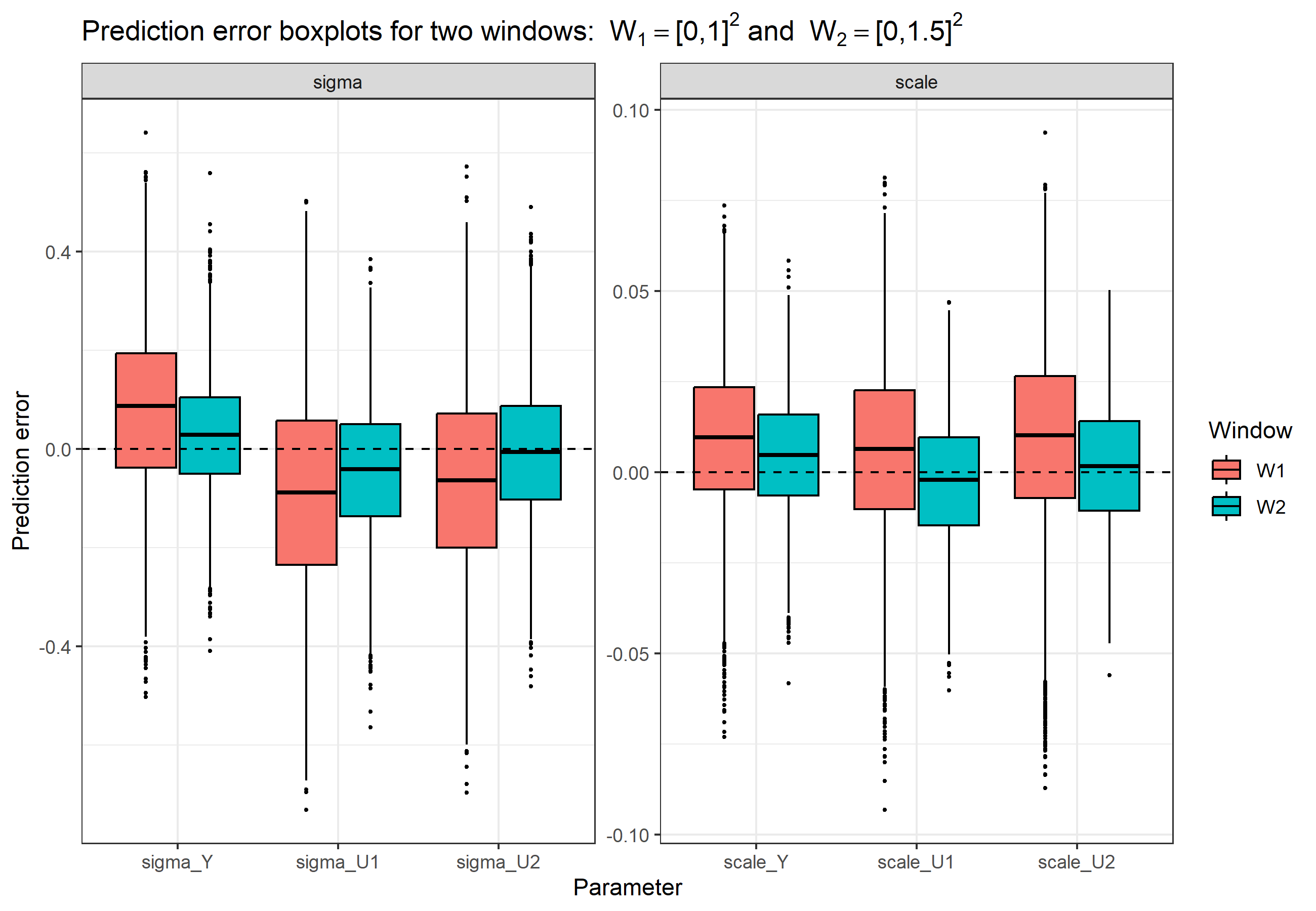}
    \caption{Prediction error results obtained using \texttt{DSBI} with two window sizes.}
    \label{fig:BLGCP_DSBI}
\end{figure}

Based on Table~\ref{tab:comparison_table} and Figure~\ref{fig:BLGCP_DSBI}, the proposed \texttt{DSBI} method performed reasonably well, with better performance when the observation window was larger. For all six second-order parameters, both RMSE and MAPE were smaller for $W_2=[0,1.5]^2$ than for $W_1=[0,1]^2$. The improvement was observed for both the variance and scale parameters. For both types of parameters, RMSE and MAPE decreased as the window size increased. This indicated that increasing the observation window improved estimation accuracy for both types of second-order parameters. These reductions showed that the larger window provided more spatial information, leading to more accurate and stable estimation of both variance and scale parameters.

For $W_1$, the variance parameters showed moderate biases, although the biases were distributed around zero across the three variance parameters. This indicated that estimating the individual latent variance parameters was challenging, as the shared and species-specific latent fields were partially confounded and their contributions to the observed spatial variation were difficult to distinguish. The scale parameters tended to be overestimated when the observation window was relatively small. The mean trend parameter estimation results for $W_1$ are reported in Appendix~\ref{sec:beta_sim2}. 

When the window size increased to $W_2$, the bias values moved closer to zero for most parameters. In particular, the biases for $\sigma_Y$, $\sigma_{U_1}$, and $\sigma_{U_2}$ were reduced substantially, suggesting that the estimates became closer to being unbiased as the observation window increased. This was consistent with the expected asymptotic behaviour: as the spatial window became larger, more information about the underlying latent fields was observed, and the estimator was expected to become approximately unbiased. The boxplots in Figure~\ref{fig:BLGCP_DSBI} further supported this finding, showing that the prediction errors under $W_2$ were more tightly concentrated around zero and showed smaller variability than those under $W_1$. These results demonstrated that \texttt{DSBI} was able to provide reasonable estimates even for relatively small observation windows and its performance improved markedly with increasing window size, yielding more stable, less biased and more precise parameter estimates. 

\section{Application}
\begin{figure}[H]
    \centering
    \includegraphics[width=0.4\linewidth]{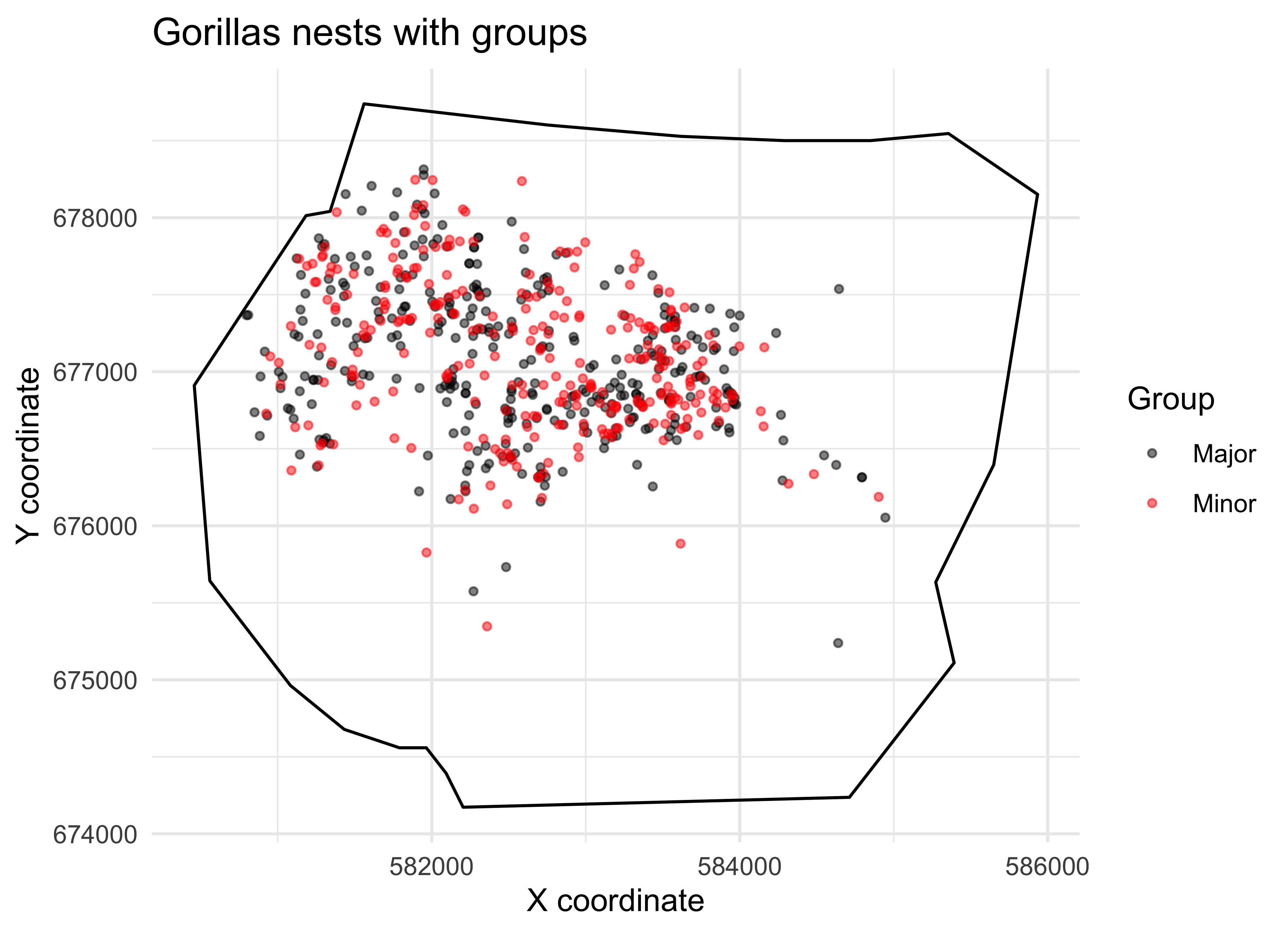}
    \caption{Locations of major group (black) and minor group (red) of the gorilla dataset.}
    \label{fig:gordata}
\end{figure}

To illustrate the practical applicability of the proposed method, we applied \texttt{DSBI} to the gorilla nest dataset available in the \textbf{spatstat} package \citep{baddeley2005spatstat}. The dataset recorded the locations of gorilla nests in the Kagwene Gorilla Sanctuary, Cameroon and was widely used as a benchmark for spatial point process modelling \citep{funwi2012understanding}. The gorilla dataset contained $647$ points in a window with a rectangular polygonal boundary $[580457.9, 585934] \times [674172.8, 678739.2]$ meters, of which $350$ belonged to the major group and $297$ belonged to the minor group. Following \cite{dovers2024fitting}, we used three continuous covariates after centring and scaling them: elevation (\texttt{elevation}), slope angle (\texttt{slopeangle}), and distance to the nearest water (\texttt{waterdist}), and three categorical covariates: average heat level (\texttt{heat}), slope type (\texttt{slopetype}), and vegetation type (\texttt{vegetation}) to fit the mean trend of the LGCP model. The gorilla dataset contained two groups, major and minor, as shown in Figure~\ref{fig:gordata}. It was clear from the visualisation that there existed a positive correlation between the two groups.

\begin{figure}[ht]
    \centering
    \includegraphics[width=0.32\linewidth]{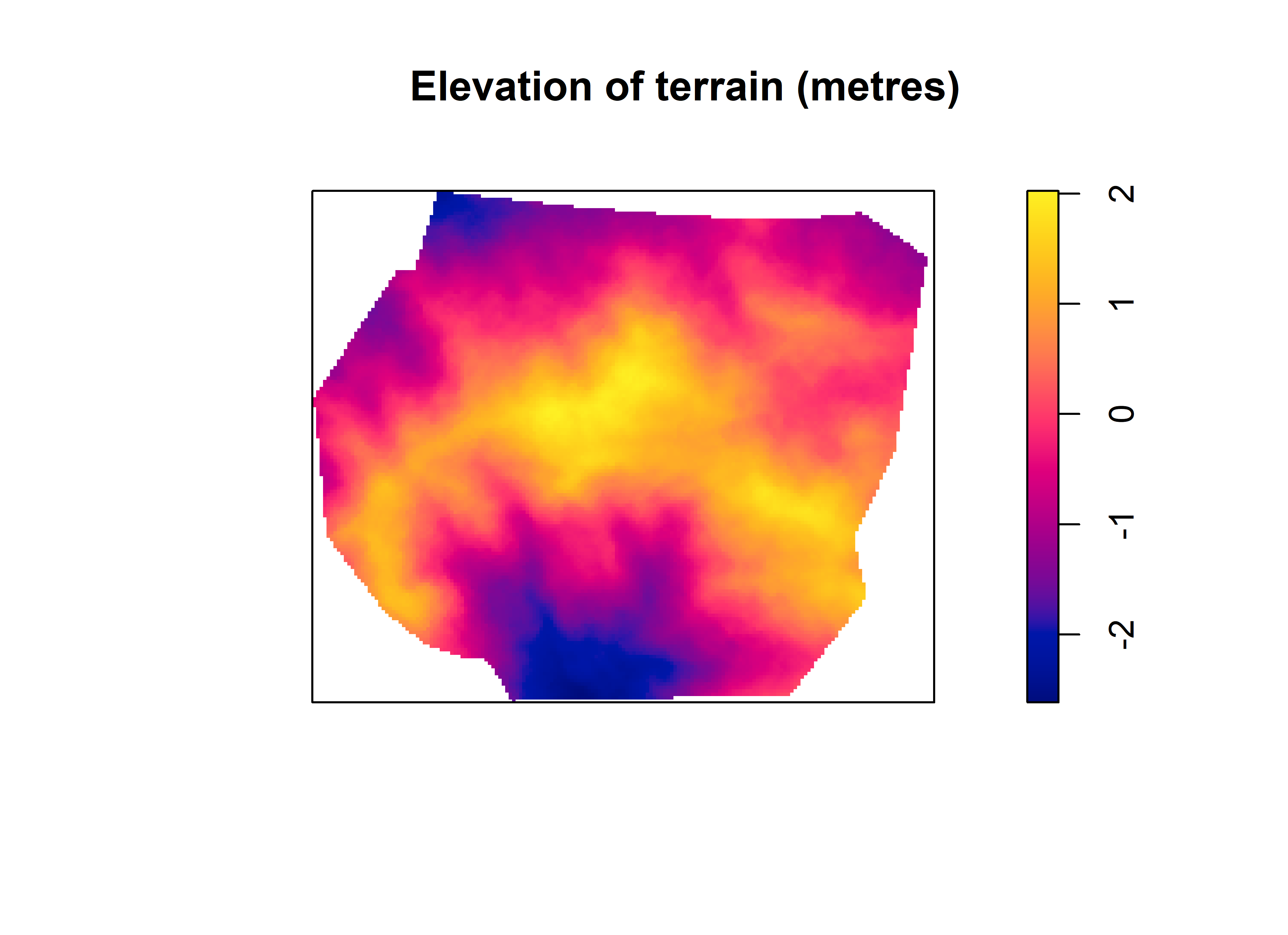}
    \hfill
    \includegraphics[width=0.32\linewidth]{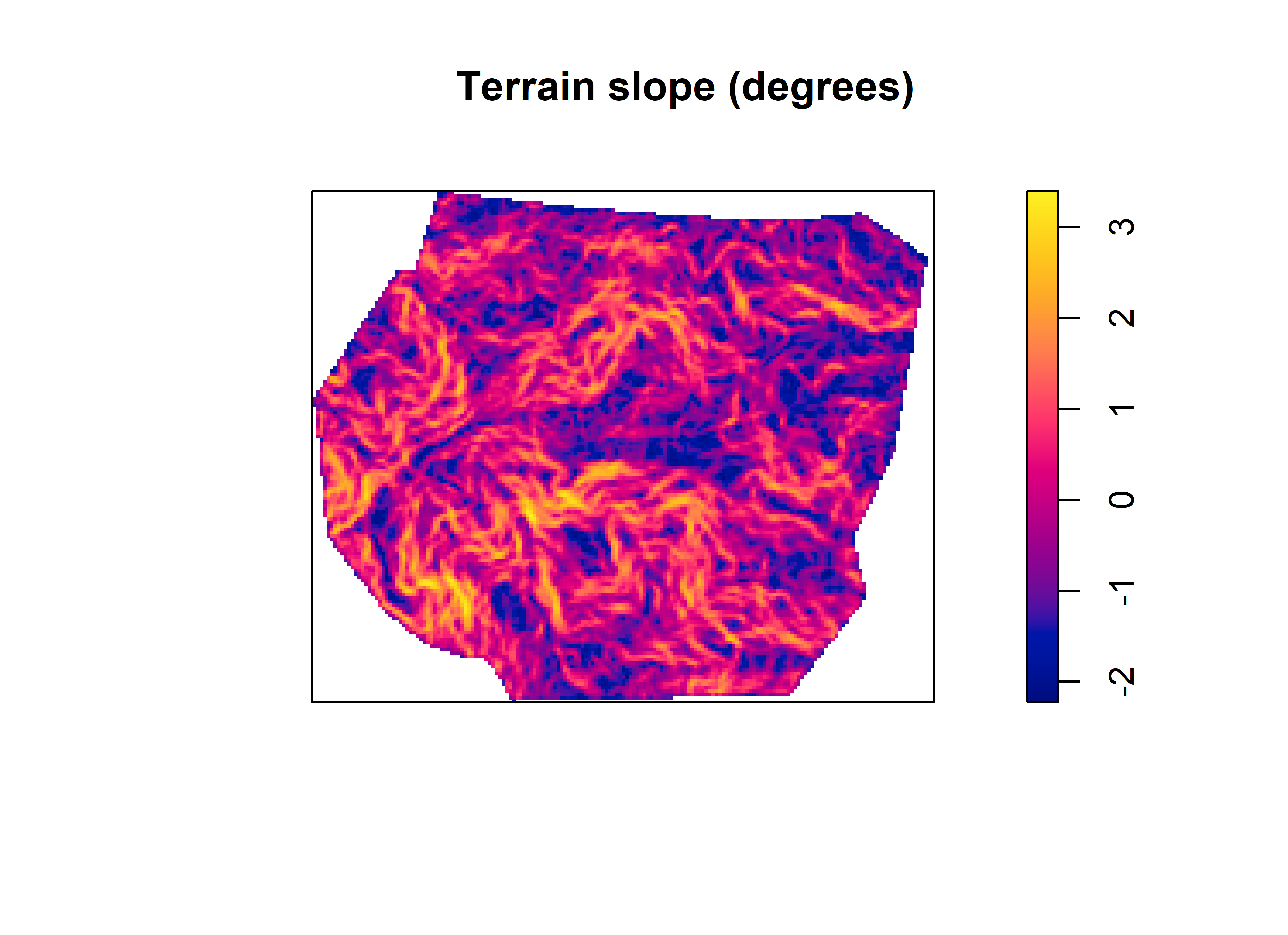}
    \hfill
    \includegraphics[width=0.32\linewidth]{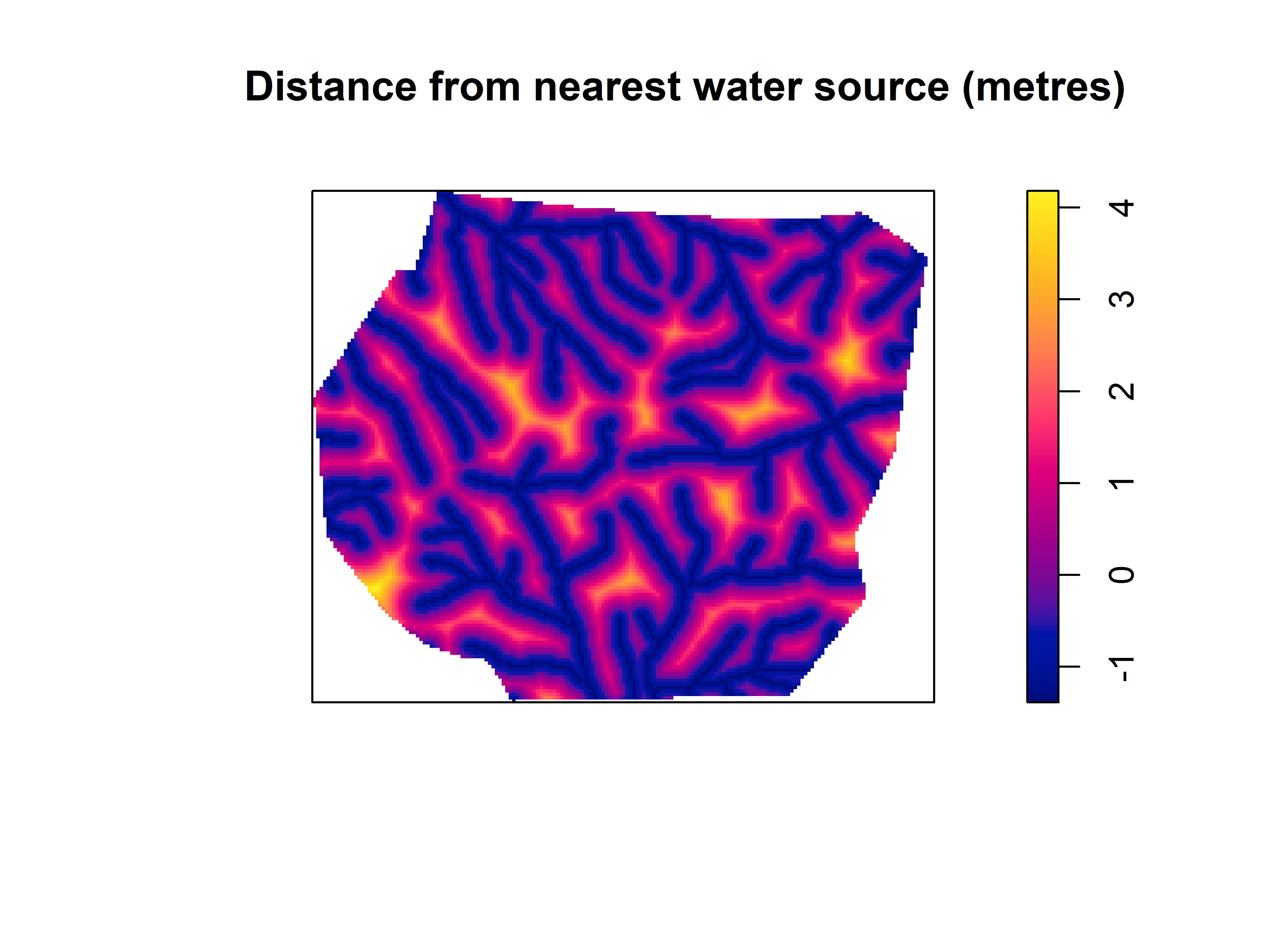}
    \hfill
    \includegraphics[width=0.32\linewidth]
    {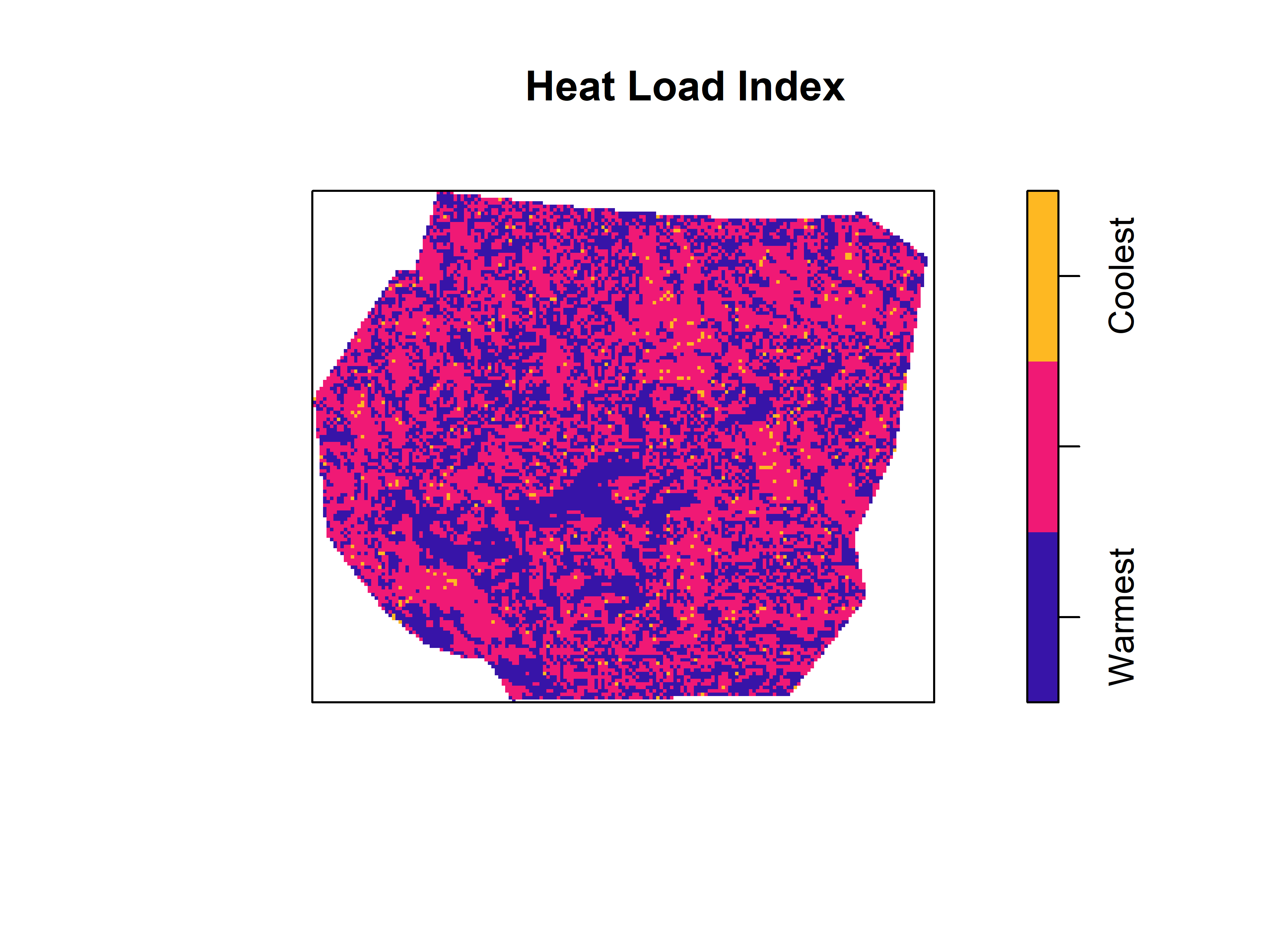}
    \hfill
    \includegraphics[width=0.32\linewidth]{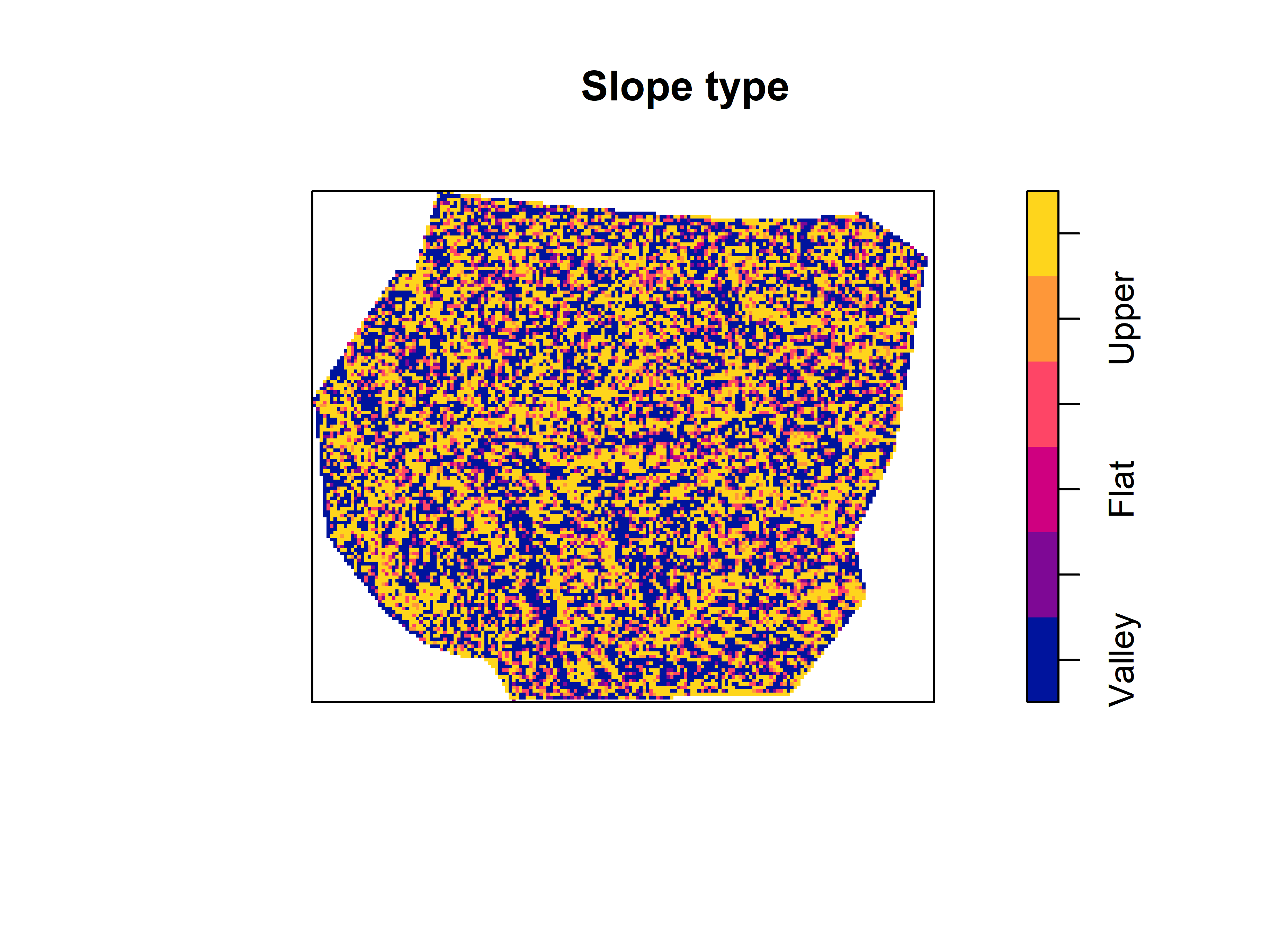}
    \hfill
    \includegraphics[width=0.32\linewidth]{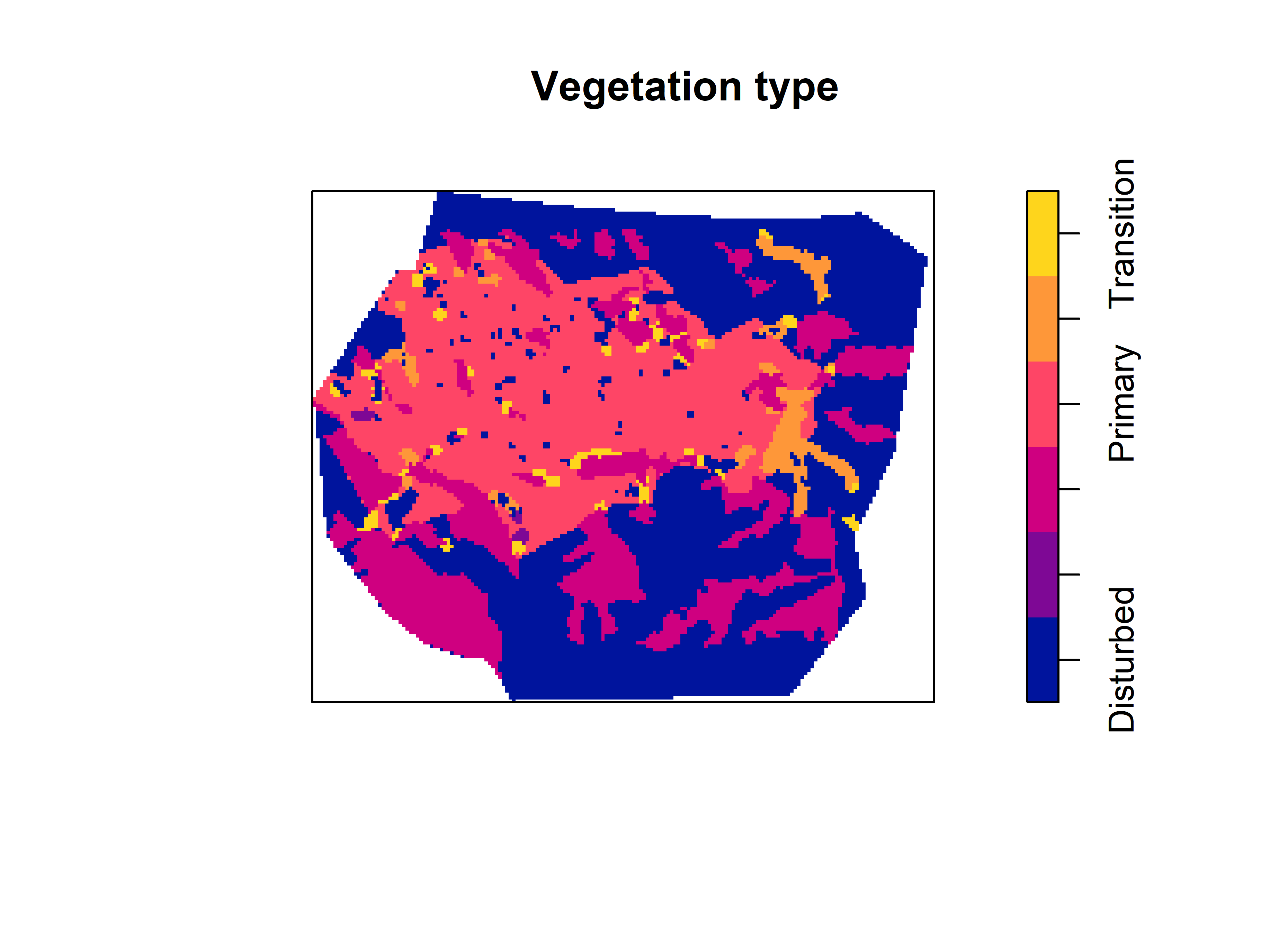}
    \caption{Covariates include elevation, slope angle, distance to water, heat, slope type, and vegetation type. Continuous covariates are presented in this figure prior to centring and scaling.}
    \label{fig:gorillasplot}
\end{figure}

The sampling ranges of $\bm{\beta}$ were chosen from the 95\% confidence intervals of the estimates obtained using \texttt{ppm()}. For the minor group, all six covariates were included. For the major group, the categorical covariate, slope type, was omitted because it produced unstable coefficient estimates with very wide confidence intervals and showed no evidence of a significant effect. The sampling ranges of the scale parameters for the training data were determined from the observed minimum and maximum interpoint distances in the combined point pattern and in the two individual point patterns $X_{\text{major}}$ and $X_{\text{minor}}$. The resulting sampling ranges were $\xi_Y \in (0.842, 869.881)$, $\xi_{U_1} \in (1.563, 869.881)$ and $\xi_{U_2} \in (2.478, 679.928)$.

Each of $\sigma_Y$, $\sigma_{U_1}$ and $\sigma_{U_2}$ was sampled from $[0.1, 1.5]$. Thus, for each type $p \in {1,2}$, the total variance $\sigma_Y^2 + \sigma_{U_p}^2$ ranged up to $1.5^2 + 1.5^2 = 4.5$. We considered a total variance of up to $4.5$ to be sufficiently large for real-world situations, while remaining computationally feasible.

\begin{figure}[H]
    \centering
    \includegraphics[width=1\linewidth]{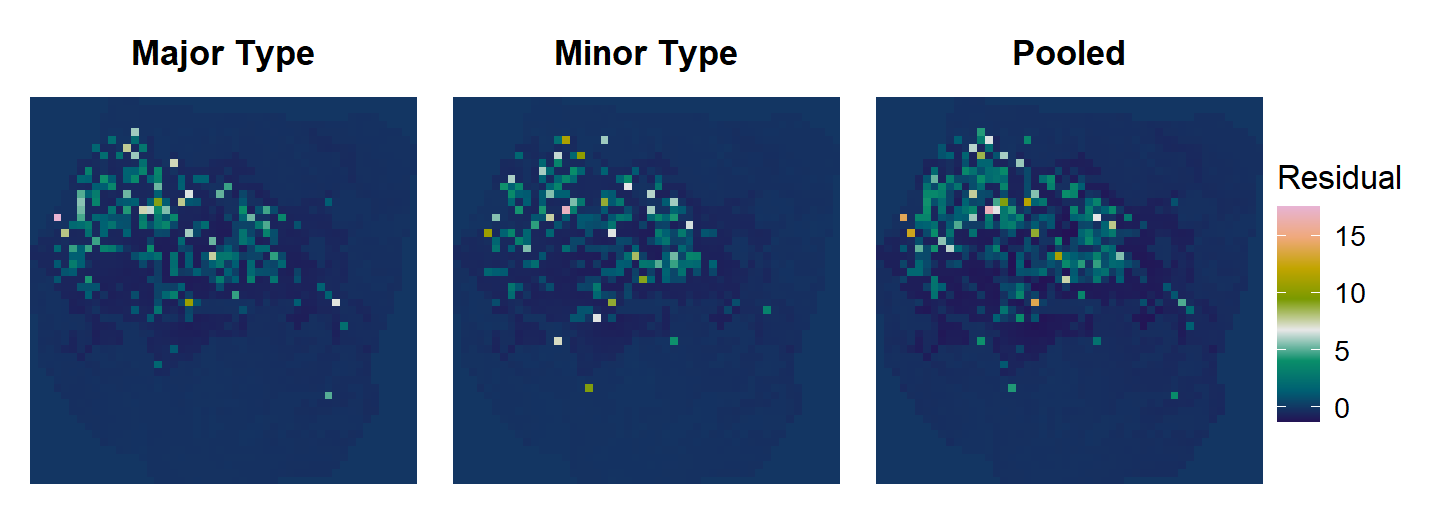}
    \caption{Spatial standardised count images for the gorilla observation, obtained by subtracting the expected counts under the fitted first-order intensity model from the observed counts: Major, Minor and Pooled (both Major and Minor).}
    \label{fig:app_count_images}
\end{figure}

We used three types of inputs $S = {S_1, S_2, S_3}$ into our model, i.e., $S_1 = \{ \hat{\bm{\beta}}_{\text{major}}, \hat{\bm{\beta}}_{\text{minor}}\}$, second-order summary statistics $S_2 = \{L_{\text{major}}- L_{\text{cross}}, L_{\text{minor}} - L_{\text{cross}}, L_{\text{cross}} - r\}$ and the standardised count images of the major type, minor type, and pooled types on a $50 \times 50$ grid. The observed standardised count images for the gorilla dataset were shown in Figure~\ref{fig:app_count_images}. Positive residuals in the images suggested clustering that was not captured by the Poisson mean trend. As in the simulation studies, we applied a square-root transformation to the response inputs and then rescaled the estimates by squaring them to enforce positivity of the latent field parameters.
 
\begin{table}[H]
\centering
\begin{tabular}{cccccc}
\toprule
\multicolumn{2}{c}{Shared} & \multicolumn{2}{c}{Major} & \multicolumn{2}{c}{Minor} \\
$\hat{\sigma}_Y$ & $\hat{\xi}_Y$ 
& $\hat{\sigma}_{U_1}$ & $\hat{\xi}_{U_1}$ 
& $\hat{\sigma}_{U_2}$ & $\hat{\xi}_{U_2}$ \\
\midrule
1.143&710.099&1.283&371.697&0.730&283.127 \\
\bottomrule
\end{tabular}
\caption{Estimated latent field parameters of the BLGCP model based on 10 training runs with different random seeds (30 epochs each). Reported values are the averages across runs.}
\label{tab:mlgcp_app_estparams}
\end{table}

\begin{figure}[H]
    \centering
    \includegraphics[width=0.32\linewidth]{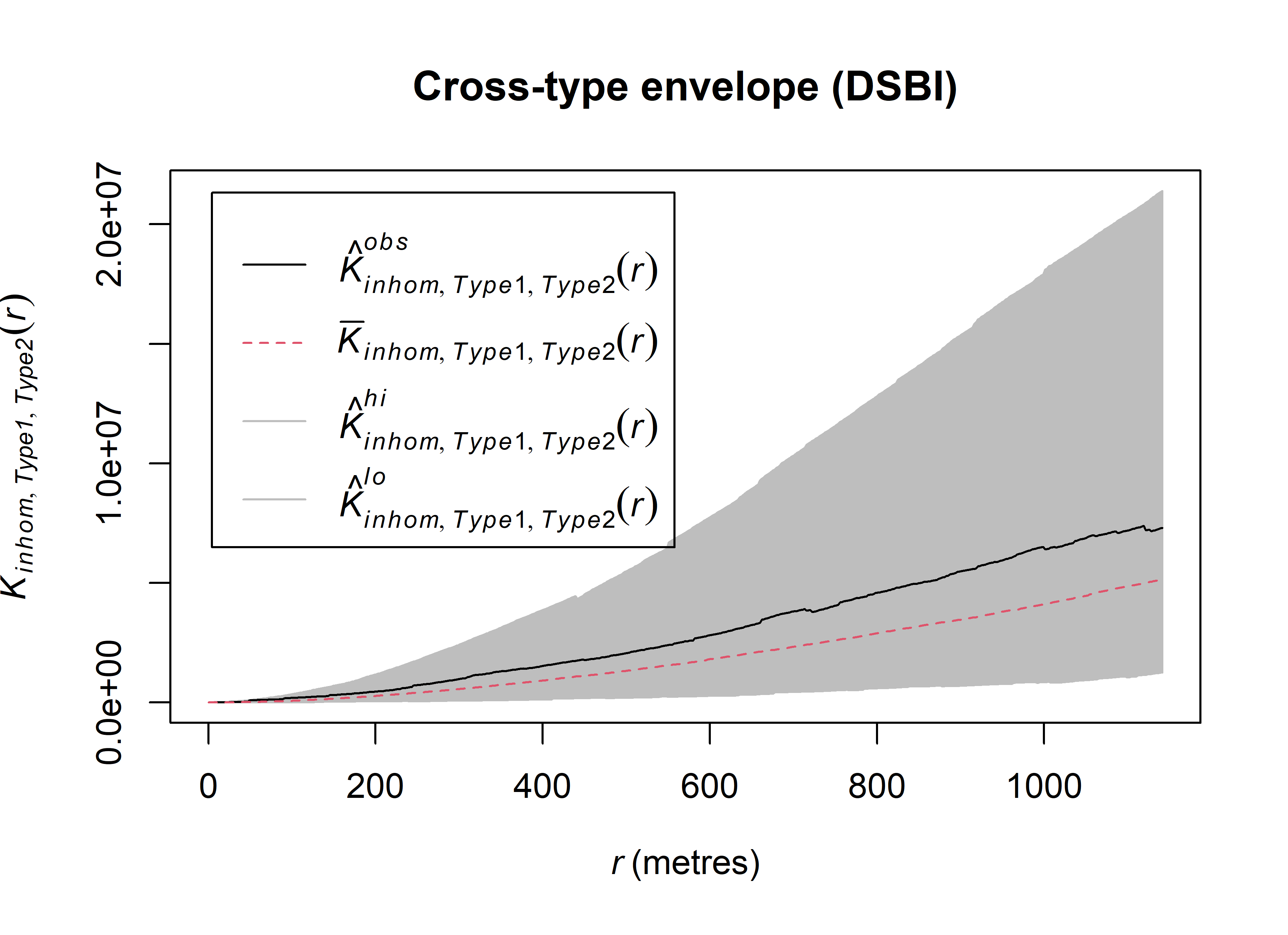}
    \includegraphics[width=0.32\linewidth]{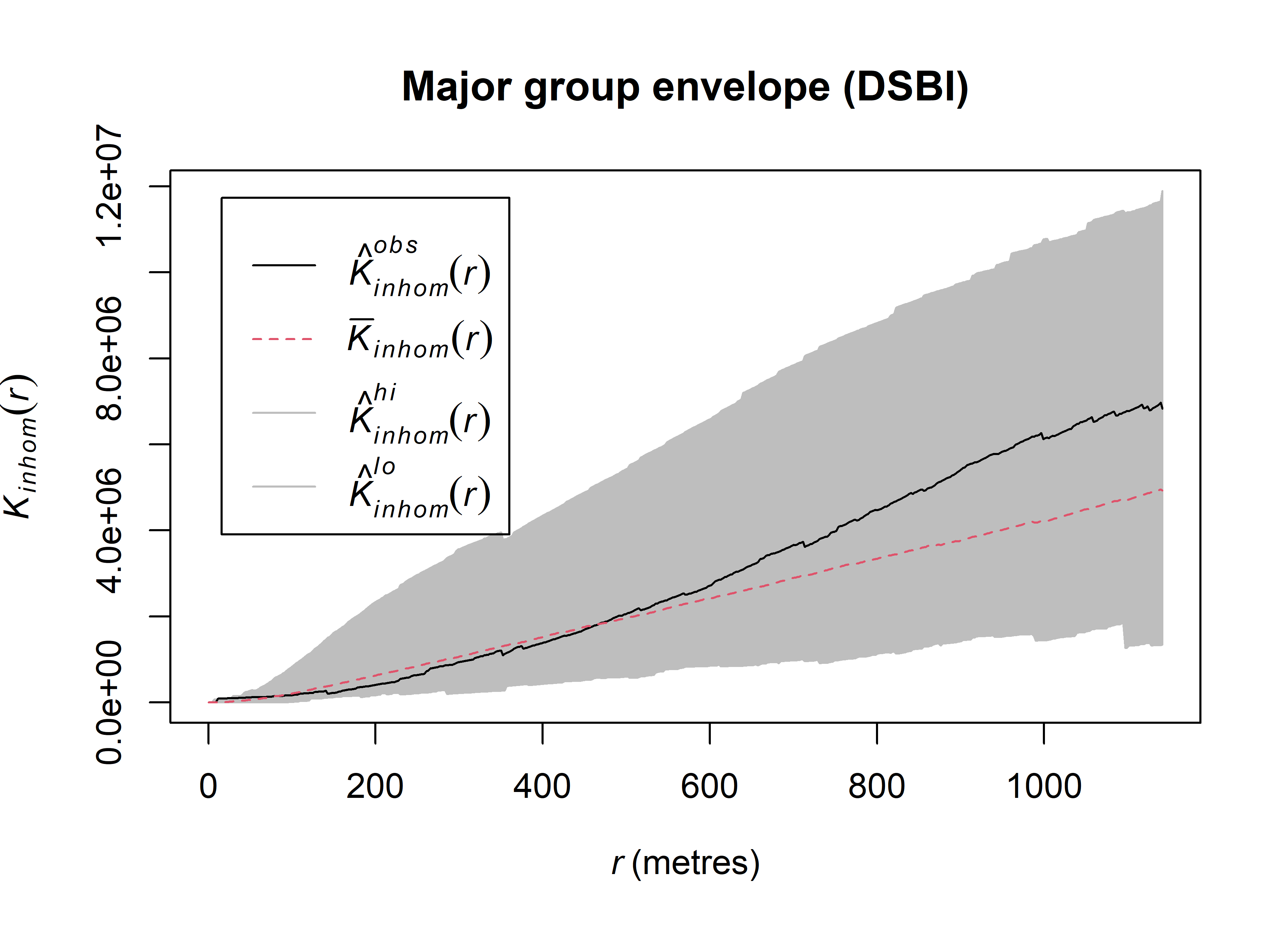}
    \includegraphics[width=0.32\linewidth]{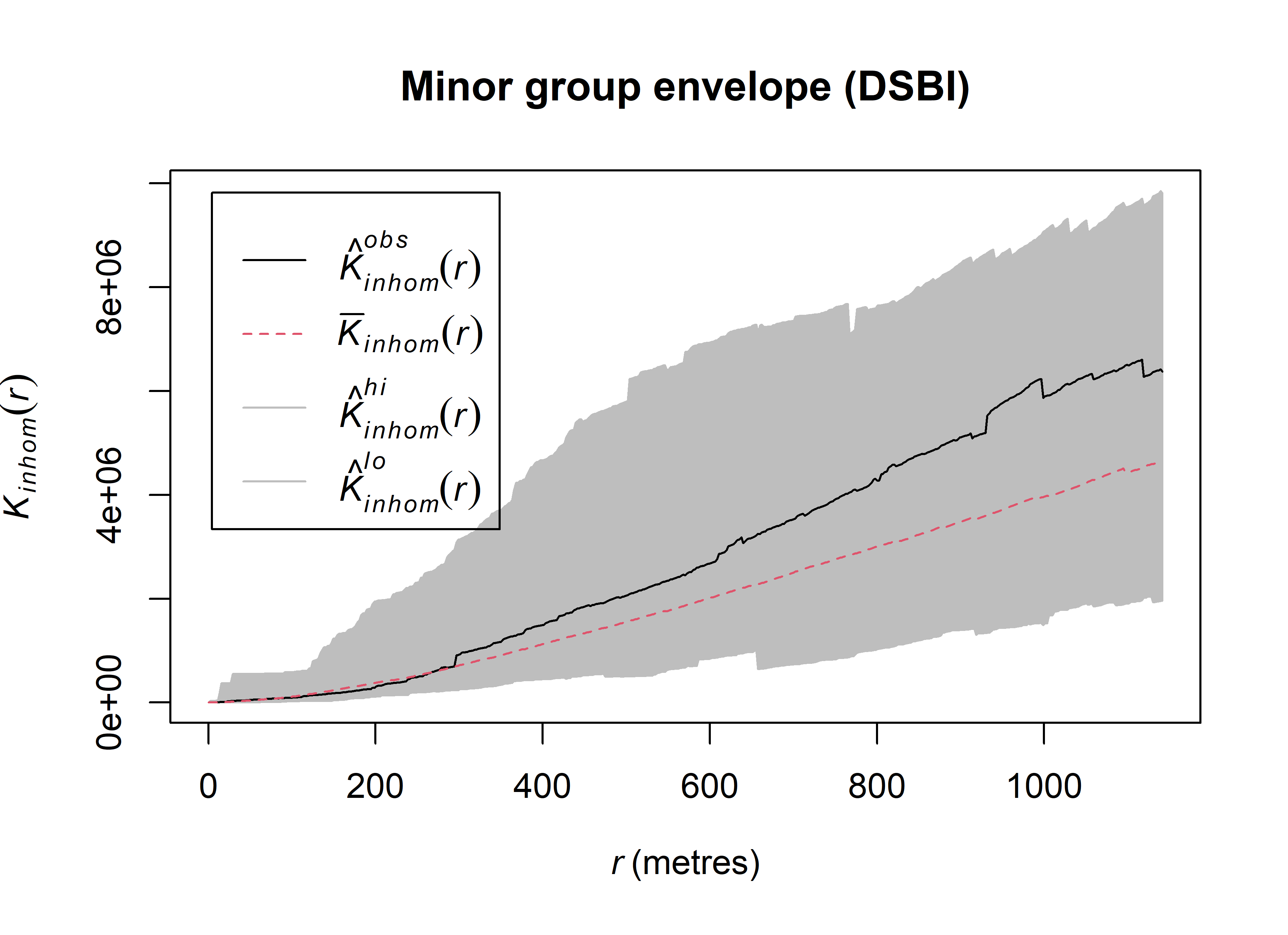}
    \caption{The p-value for cross type is 0.38, and the p-values for major type is 0.41, and minor type are 0.15.}
    \label{fig:mlgcp_env}
\end{figure}

Table~\ref{tab:mlgcp_app_estparams} showed the estimated latent field parameters of the shared field and individual fields. The fitted shared latent field had an estimated standard deviation of
$\hat\sigma_Y=1.143$ and range parameter
$\hat\xi_Y=710.099$. This indicated a substantial amount of large-scale spatial variation that was common to both gorilla groups after accounting for the deterministic mean trend. The relatively large estimated range suggested that the shared environmental effects were spatially correlated over long distances. The major-group-specific latent field had the largest variability $\hat\sigma =1.283$, indicating stronger residual clustering unique to the major group than to the minor group. Its estimated range $\hat\xi_{U_1}=371.697$ was smaller than that of the shared field, suggesting that this additional clustering occurred at a more local spatial scale. In contrast, the minor-group-specific field had a smaller variability $\hat\sigma_{U_2}=0.730$, implying weaker residual clustering after accounting for both the shared field and the deterministic trend. Its estimated range $\hat\xi_{U_2}=283.127$ was the smallest of the three latent fields, indicating that the remaining group-specific dependence was relatively local. The computational time and comparison with the estimation results from INLA using the two-step procedure were reported in Appendix~\ref{sec:appendix_time}. In addition, we provided examples of realisations of simulated gorilla locations using the fitted bivariate LGCP model with the estimated parameters shown in Table~\ref{tab:mlgcp_app_estparams} in Appendix~\ref{sec:app_sim_gorilla}.

We used simulation envelopes to validate the fit of the methods. Specifically, we simulated bivariate LGCP realisations based on the fitted parameters $\hat \sigma^2$ and $\hat \xi$ and used the simulated envelopes to assess model adequacy. As a supplement to the envelope plots, we also used the Diggle-Cressie-Loosmore-Ford (DCLF) test implemented in \texttt{spatstat} \citep{baddeley2014tests}. The DCLF test was a Monte Carlo goodness-of-fit test based on the discrepancy between the observed summary function and the summary functions simulated under a fitted model. We used the estimated mean trend and covariance parameters $(\hat \sigma^{2}, \hat \xi)$ to simulate BLGCP processes for constructing the envelopes.

In this work, the null hypothesis was that the observed gorilla point pattern was a realisation of a bivariate LGCP with the estimated deterministic mean trend and covariance parameters $(\hat \sigma^{2}, \hat \xi)$. That is, $$ H_0: \text{the observed point pattern is generated from the fitted model}. $$ The alternative hypothesis was that the fitted bivariate LGCP model did not adequately describe the observed point pattern: $$ H_1: \text{the observed point pattern is not generated from the fitted model}.$$ Under $H_0$, the simulated bivariate LGCP realisations were treated as samples from the fitted model and were used to construct the simulation envelopes and compute the DCLF test statistic. We set the significance level to $0.05$. Therefore, if the p-value was greater than $0.05$, we failed to reject $H_0$, suggesting that the estimated parameters $(\hat \sigma^{2}, \hat \xi)$ adequately reproduced the observed clustering structure. Conversely, if the p-value was less than $0.05$, we rejected $H_0$, suggesting that the fitted LGCP model with parameters $(\hat \sigma^{2}, \hat \xi)$ did not adequately match the observed clustering structure and may have been misspecified.

Figure~\ref{fig:mlgcp_env} showed a wide envelope, which was likely attributable to the relatively large estimated latent field variance, $\hat\sigma^2$, that generated considerable between realisation variability. This suggested that there may still have been unknown effects or covariates that had not been identified and were therefore captured by the latent fields. Nevertheless, the bivariate LGCP model provided an adequate fit to the observed data, since all observed K-functions remained within the envelope. Based on $99$ simulations, the p-value for the cross type was $0.38$, and the p-values for the major and minor types were $0.41$ and $0.15$ respectively, with no significant deviation detected at the $5\%$ level. These results indicated that the fitted bivariate LGCP adequately captured both the individual-type and cross-type spatial dependence structures present in the observed data.

\section{Discussion}
Building on the two-step estimation procedure of \cite{waagepetersen2007estimating} and \cite{waagepetersen2009two}, we propose a two-step deep simulation-based inference approach for bivariate LGCP. This approach integrates the robustness of classical statistical methods for estimating mean trend parameters with the flexibility of simulation-based approaches for estimating latent field parameters. This work extends the homogeneous simulation-based inference framework proposed by \cite{vihrs2022using} to a more general framework that accommodates inhomogeneous and multi-type settings. 

To address the challenges in estimating latent field parameters, we introduce a new class of summary statistics, referred to as spatial structure inputs. These are two-dimensional images designed to directly capture the spatial structure of the observed data. It allows the model to learn directly from spatial inputs and complements classical summary statistics, thereby enhancing the incorporation of spatial information in the learning process. The two-step estimation procedure introduced in this work for deep simulation-based inference separates the estimation of first-order and second-order components. This approach reduces the number of parameters that the model needs to learn. In particular, only the latent field parameters are learned, while the mean trend parameters are estimated reliably using a classical Poisson model. The hybrid estimation strategy reduces computational cost and removes the need to specify broad simulation ranges for mean trend parameters when covariates are present. As the model extends from univariate to multivariate settings, the number of both mean trend and latent field parameters increases substantially. By decoupling their estimation, the proposed approach allows the model to focus on the second-order latent field parameters, thereby reducing inherent identifiability issues in multivariate processes. In addition, once the \texttt{DSBI} model is pre-trained, it provides much faster estimation than classical likelihood-based methods, as the simulation-based inference requires only a simple mapping. This allows the pre-trained model to be reused efficiently: if new observed data are similar to the training data, the model can be applied directly to obtain results without any optimisation, unlike traditional methods. This is particularly useful for repeated studies with similar data or for analyses involving point patterns with similar underlying structures. Although we illustrate the method for inhomogeneous Bivariate LGCPs, it is not limited to this class. Because the approach relies only on simulation and the separation of first-order and second-order components, it can be applied to a wide range of spatial point process models, for which likelihood-based inference is often computationally demanding or intractable. In this work, we apply the proposed method to the gorilla dataset, which consists of two types of point patterns, to demonstrate its applicability in a real-world setting. The \texttt{DSBI} method provides a good fit to the data according to envelope tests. However, it produces a slightly larger variance parameter, indicating that the available covariates may be incomplete and that some relevant covariates remain unobserved.

Future work includes extending the method to multivariate point processes with more types, where estimating latent field parameters becomes challenging as the number of latent fields increases and purely statistical methods become impractical. In addition, the class of spatial structure inputs can be further explored. At present, we consider only count images. However, alternative representations, such as density images, residual images, and their combinations, may provide additional useful information. Moreover, the design of the neural network architecture needs further investigation. The simple convolutional network employed in this work may be insufficient to capture the complex dependencies present in multi-type point processes, resulting in biased predicted values in scale parameters. More advanced architectures, such as ResNet or other modern convolutional neural network designs, may offer improved performance. An important open question is how to define appropriate sampling ranges for model parameters. Although the proposed method decouples the estimation of mean trend and latent field parameters, it still relies on predefined ranges. While this is tractable in univariate settings, multivariate models introduce a larger number of latent fields and more complex parameter dependencies. Resampling-based approaches, including bootstrap, Monte Carlo, and Bayesian methods, may provide a principled way to determine suitable sampling ranges. Another possible direction is to pre-train a large-scale neural network, analogous to Generative Pre-trained Transformer models, on a collection of simulated point patterns generated from a wide range of underlying processes. By including rich summary statistics, such a model could learn general representations of spatial structures across different models. Once trained, this model could be applied directly to new observed point patterns without requiring additional optimisation or retraining. Given an observed dataset, the model would infer the most plausible underlying point process structure and provide corresponding parameter estimates immediately. Such a model would significantly reduce computational costs and enable rapid inference in repeated studies.

\section*{Acknowledgements}
This research was supported by the Commonwealth through an Australian Government Research Training Program Scholarship [DOI: https://doi.org/10.82133/C42F-K220].

\section*{Conflicts of Interest}
The authors declare no conflicts of interest.

\bibliographystyle{apalike}
\bibliography{B}

@article{agarap2018deep,
  title={Deep learning using rectified linear units (relu)},
  author={Agarap, Abien Fred},
  journal={arXiv preprint arXiv:1803.08375},
  year={2018}
}

@article{baddeley2000non,
  title={{N}on-and semi-parametric estimation of interaction in inhomogeneous point patterns},
  author={Baddeley, Adrian J and M{\o}ller, Jesper and Waagepetersen, Rasmus},
  journal={Statistica Neerlandica},
  volume={54},
  number={3},
  pages={329--350},
  year={2000},
  publisher={Wiley Online Library}
}

@article{brix2001spatiotemporal,
  title={{S}patiotemporal prediction for log-{G}aussian {C}ox processes},
  author={Brix, Anders and Diggle, Peter J},
  journal={Journal of the Royal Statistical Society: Series B (Statistical Methodology)},
  volume={63},
  number={4},
  pages={823--841},
  year={2001},
  publisher={Wiley Online Library}
}

@article{baddeley2005spatstat,
  title={Spatstat: an {R} package for analyzing spatial point patterns},
  author={Baddeley, Adrian and Turner, Rolf},
  journal={Journal of statistical software},
  volume={12},
  pages={1--42},
  year={2005}
}

@article{baddeley2014tests,
  title={On tests of spatial pattern based on simulation envelopes},
  author={Baddeley, Adrian and Diggle, Peter J and Hardegen, Andrew and Lawrence, Thomas and Milne, Robin K and Nair, Gopalan},
  journal={Ecological monographs},
  volume={84},
  number={3},
  pages={477--489},
  year={2014},
  publisher={Wiley Online Library}
}

@book{baddeley2016spatial,
  title={Spatial point patterns: methodology and applications with {R}},
  author={Baddeley, Adrian and Rubak, Ege and Turner, Rolf},
  volume={1},
  year={2016},
  publisher={CRC press Boca Raton}
}

@article{choiruddin2020regularized,
  title={Regularized estimation for highly multivariate log {G}aussian {C}ox processes},
  author={Choiruddin, Achmad and Cuevas-Pacheco, Francisco and Coeurjolly, Jean-Fran{\c{c}}ois and Waagepetersen, Rasmus},
  journal={Statistics and Computing},
  volume={30},
  number={3},
  pages={649--662},
  year={2020},
  publisher={Springer}
}

@article{cranmer2020frontier,
  title={The frontier of simulation-based inference},
  author={Cranmer, Kyle and Brehmer, Johann and Louppe, Gilles},
  journal={Proceedings of the National Academy of Sciences},
  volume={117},
  number={48},
  pages={30055--30062},
  year={2020},
  publisher={National Academy of Sciences}
}

@article{diggle1983bivariate,
  title={Bivariate Cox processes: some models for bivariate spatial point patterns},
  author={Diggle, Peter J and Milne, Robin K},
  journal={Journal of the Royal Statistical Society Series B: Statistical Methodology},
  volume={45},
  number={1},
  pages={11--21},
  year={1983},
  publisher={Oxford University Press}
}

@book{daley2003introduction,
  title={An introduction to the theory of point processes: volume {I}: elementary theory and methods},
  author={Daley, Daryl J and Vere-Jones, David},
  year={2003},
  publisher={Springer}
}

@article{diggle1984monte,
  title={{M}onte {C}arlo methods of inference for implicit statistical models},
  author={Diggle, Peter J and Gratton, Richard J},
  journal={Journal of the Royal Statistical Society Series B: Statistical Methodology},
  volume={46},
  number={2},
  pages={193--212},
  year={1984},
  publisher={Oxford University Press}
}

@article{dovers2024fitting,
  title={Fitting log-{G}aussian {C}ox processes using generalized additive model software},
  author={Dovers, Elliot and Stoklosa, Jakub and Warton, David I},
  journal={The American Statistician},
  volume={78},
  number={4},
  pages={418--425},
  year={2024},
  publisher={Taylor \& Francis}
}

@article{flagg2023integrated,
  title={The integrated nested {L}aplace approximation applied to spatial log-{G}aussian {C}ox process models},
  author={Flagg, Kenneth and Hoegh, Andrew},
  journal={Journal of Applied Statistics},
  volume={50},
  number={5},
  pages={1128--1151},
  year={2023},
  publisher={Taylor \& Francis}
}

@article{funwi2012understanding,
  title={Understanding the nesting spatial behaviour of gorillas in the {K}agwene {S}anctuary, {C}ameroon},
  author={Funwi-Gabga, Neba and Mateu, Jorge},
  journal={Stochastic Environmental Research and Risk Assessment},
  volume={26},
  number={6},
  pages={793--811},
  year={2012},
  publisher={Springer}
}

@article{guan2006composite,
  title={A composite likelihood approach in fitting spatial point process models},
  author={Guan, Yongtao},
  journal={Journal of the American Statistical Association},
  volume={101},
  number={476},
  pages={1502--1512},
  year={2006},
  publisher={Taylor \& Francis}
}

@inproceedings{lueckmann2021benchmarking,
  title={Benchmarking simulation-based inference},
  author={Lueckmann, Jan-Matthis and Boelts, Jan and Greenberg, David and Goncalves, Pedro and Macke, Jakob},
  booktitle={International conference on artificial intelligence and statistics},
  pages={343--351},
  year={2021},
  organization={PMLR}
}

@article{lindgren2015bayesian,
  title={Bayesian spatial modelling with {R}-{I}{N}{L}{A}},
  author={Lindgren, Finn and Rue, H{\aa}vard},
  journal={Journal of statistical software},
  volume={63},
  pages={1--25},
  year={2015}
}

@article{moller1998log,
  title={Log {G}aussian {C}ox processes},
  author={M{\o}ller, Jesper and Syversveen, Anne Randi and Waagepetersen, Rasmus Plenge},
  journal={Scandinavian journal of statistics},
  volume={25},
  number={3},
  pages={451--482},
  year={1998},
  publisher={Wiley Online Library}
}

@article{mateu2022spatial,
  title={Spatial point processes and neural networks: {A} convenient couple},
  author={Mateu, Jorge and Jalilian, Abdollah},
  journal={Spatial Statistics},
  volume={50},
  pages={100644},
  year={2022},
  publisher={Elsevier}
}

@book{moraga2023spatial,
  title={Spatial statistics for data science: theory and practice with {R}},
  author={Moraga, Paula},
  year={2023},
  publisher={Chapman and Hall/CRC}
}

@article{rue2009approximate,
  title={Approximate {B}ayesian inference for latent {G}aussian models by using integrated nested {L}aplace approximations},
  author={Rue, H{\aa}vard and Martino, Sara and Chopin, Nicolas},
  journal={Journal of the Royal Statistical Society Series B: Statistical Methodology},
  volume={71},
  number={2},
  pages={319--392},
  year={2009},
  publisher={Oxford University Press}
}

@article{tanaka2008parameter,
  title={Parameter estimation and model selection for {N}eyman-{S}cott point processes},
  author={Tanaka, Ushio and Ogata, Yosihiko and Stoyan, Dietrich},
  journal={Biometrical Journal: Journal of Mathematical Methods in Biosciences},
  volume={50},
  number={1},
  pages={43--57},
  year={2008},
  publisher={Wiley Online Library}
}

@article{teng2017bayesian,
  title={Bayesian computation for {L}og-{G}aussian {C}ox processes: {A} comparative analysis of methods},
  author={Teng, Ming and Nathoo, Farouk and Johnson, Timothy D},
  journal={Journal of statistical computation and simulation},
  volume={87},
  number={11},
  pages={2227--2252},
  year={2017},
  publisher={Taylor \& Francis}
}

@article{taylor2015bayesian,
  title={Bayesian inference and data augmentation schemes for spatial, spatiotemporal and multivariate log-{G}aussian {C}ox processes in {R}},
  author={Taylor, Benjamin M and Davies, Tilman M and Rowlingson, Barry S and Diggle, Peter J},
  journal={Journal of Statistical Software},
  volume={63},
  pages={1--48},
  year={2015}
}

@article{taylor2014inla,
  title={{INLA} or {MCMC}? {A} tutorial and comparative evaluation for spatial prediction in log-{G}aussian {C}ox processes},
  author={Taylor, Benjamin M and Diggle, Peter J},
  journal={Journal of Statistical Computation and Simulation},
  volume={84},
  number={10},
  pages={2266--2284},
  year={2014},
  publisher={Taylor \& Francis}
}

@article{vihrs2022using,
  title={Using neural networks to estimate parameters in spatial point process models},
  author={Vihrs, Ninna},
  journal={Spatial Statistics},
  volume={51},
  pages={100668},
  year={2022},
  publisher={Elsevier}
}

@article{waagepetersen2007estimating,
  title={An estimating function approach to inference for inhomogeneous {N}eyman--{S}cott processes},
  author={Waagepetersen, Rasmus Plenge},
  journal={Biometrics},
  volume={63},
  number={1},
  pages={252--258},
  year={2007},
  publisher={Oxford University Press}
}

@article{waagepetersen2009two,
  title={Two-step estimation for inhomogeneous spatial point processes},
  author={Waagepetersen, Rasmus and Guan, Yongtao},
  journal={Journal of the Royal Statistical Society Series B: Statistical Methodology},
  volume={71},
  number={3},
  pages={685--702},
  year={2009},
  publisher={Oxford University Press}
}

@article{waagepetersen2016analysis,
  title={Analysis of multispecies point patterns by using multivariate log-{G}aussian {C}ox processes},
  author={Waagepetersen, Rasmus and Guan, Yongtao and Jalilian, Abdollah and Mateu, Jorge},
  journal={Journal of the Royal Statistical Society Series C: Applied Statistics},
  volume={65},
  number={1},
  pages={77--96},
  year={2016},
  publisher={Oxford University Press}
}

@article{zammit2025neural,
  title={Neural methods for amortized inference},
  author={Zammit-Mangion, Andrew and Sainsbury-Dale, Matthew and Huser, Rapha{\"e}l},
  journal={Annual Review of Statistics and Its Application},
  volume={12},
  number={1},
  pages={311--335},
  year={2025},
  publisher={Annual Reviews}
}

@article{zhu2025minimum,
  title={On minimum contrast method for multivariate spatial point processes},
  author={Zhu, Lin and Yang, Junho and Jun, Mikyoung and Cook, Scott},
  journal={Electronic Journal of Statistics},
  volume={19},
  number={1},
  pages={1889--1941},
  year={2025},
  publisher={The Institute of Mathematical Statistics and the Bernoulli Society}
}

\appendix
\section{Computational Time Across Methods}
\label{sec:appendix_time}
The main hardware used in this work was as follows: Intel Core Ultra 9 285K CPU; 64 GB RAM; NVIDIA GeForce RTX 5080 16 GB GPU.

The training time for Simulation Study of homogeneous bivariate LGCP was conducted on Window 1, $[0,1]^2$, using $23$-worker parallel computing on an Intel Core Ultra 9 285K CPU are reported here. For $100000$ training samples, the computation time was $113.100$ minutes, with an additional $5.713$ minutes required to save the training data. Monte Carlo tuning of the MC method with $300$ Monte Carlo samples was performed using $23$-worker parallel computing on an Intel Core Ultra 9 285K CPU, and required $44.114$ minutes. Based on an NVIDIA GeForce RTX 5080 16 GB GPU, the proposed method took $56.090$ minutes to train $10$ neural networks with different random seeds using $100000$ training samples. For the testing data, we compared the estimation speed across the three methods. For a fair comparison, we used a single testing data without parallel computing. The dataset used was the first dataset among the $500$ generated testing datasets. See Table~\ref{tab:time_table}. 

Although the simulation and training stages are computationally intensive, they are performed only once offline. Thereafter, inference for a new dataset using the pre-trained DSBI model takes less time, making DSBI substantially faster for repeated analyses.

\begin{table}[H]
    \centering
    \begin{tabular}{cccc}
        Time &  Proposed & MC& INLA\\
        \hline
        Sampling training data & $113.100$ minutes & - & -\\
        Saving training data & $5.713$ minutes &-&-\\
        Training neural network & $56.090$ minutes & -& -\\
        Monte Carlo tuning & - & $44.114$ minutes &-\\
        Prediction on one testing data & $0.056$ minutes & $0.821$ minutes & $1.280$ minutes.\\
        \hline
    \end{tabular}
    \caption{Computation time for Simulation Study of homogeneous bivariate LGCP on Window 1, $W_1 = [0,1]^2$. Prediction times are reported for one testing dataset without parallel computing.}
    \label{tab:time_table}
\end{table}

For the application, we also included the computational times of DSBI and INLA for comparison. Simulating $10000$ training datasets for the gorilla dataset took $209.100$ minutes. Training $10$ neural networks took $5.778$ minutes. Estimation using the pre-trained model took $0.429$ seconds. The INLA estimates of the latent parameters using the two-step method are reported here. The two-step method for INLA is used to significantly reduce computational time. The estimates are $3482.651$ for $\xi_Y$, $635.681$ for $\xi_{U_1}$, $700.360$ for $\xi_{U_2}$, $3.092$ for $\sigma_Y$, $0.114$ for $\sigma_{U_1}$, and $0.112$ for $\sigma_{U_2}$. The estimated scale parameters are much larger than those obtained using the proposed method, as observed in simulation study of homogeneous bivariate LGCP. The total variance of the shared and individual fields is also slightly larger than the total variance estimated by the proposed method. For INLA with the two-step method, where INLA is used only to estimate the latent parameters, estimation took $1.105$ minutes. This confirms that the proposed method is very efficient after training.

\section{First-Order Trend Parameter Estimation for the Inhomogeneous Bivariate LGCP Simulation Study}
\label{sec:beta_sim2}
\begin{figure}[H]
    \centering
    \includegraphics[width=1\linewidth]{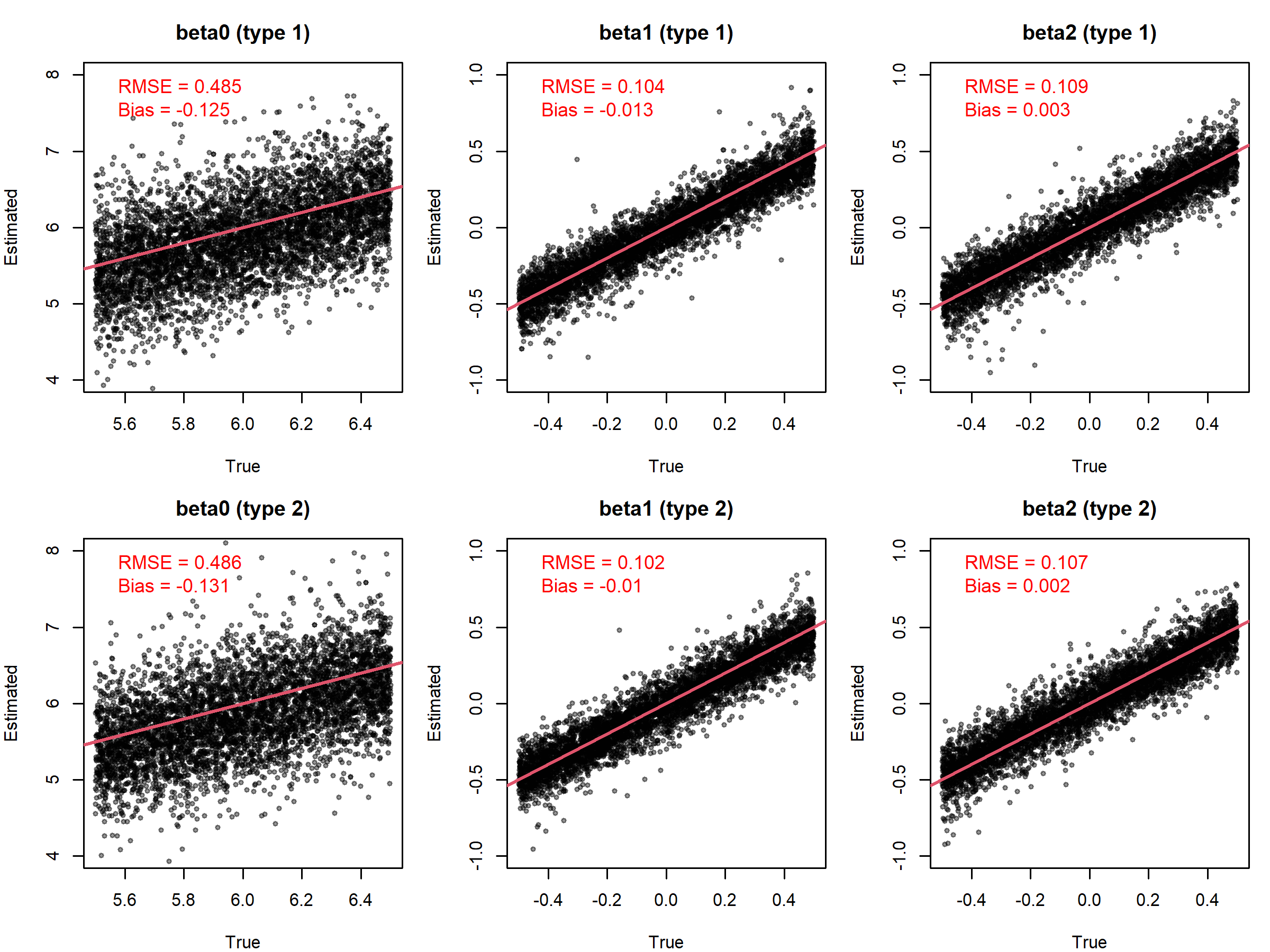}
    \caption{Estimation results for the mean trend parameters in the simulation study for the inhomogeneous bivariate LGCP with window $W_1 = [0,1]^2$. Each panel shows the estimated and true values of one regression parameter for the two types.}
    \label{fig:inhom_beta}
\end{figure}

We reported the first-order trend parameter estimates from the simulation study for the inhomogeneous bivariate LGCP in Section~\ref{sec:Inhomogeneous Bivariate LGCP} to evaluate the first-order performance of the two-step estimation procedure. Figure~\ref{fig:inhom_beta} showed that the estimates were approximately unbiased for the mean trend parameters, suggesting that the two-step method \citep{waagepetersen2007estimating, waagepetersen2009two}, with a fitted Poisson mean trend, was reliable.

\section{Simulated Realisations Using the Fitted Model for Gorilla Locations}
\label{sec:app_sim_gorilla}
\begin{figure}[H]
    \centering
    \includegraphics[width=0.48\linewidth]{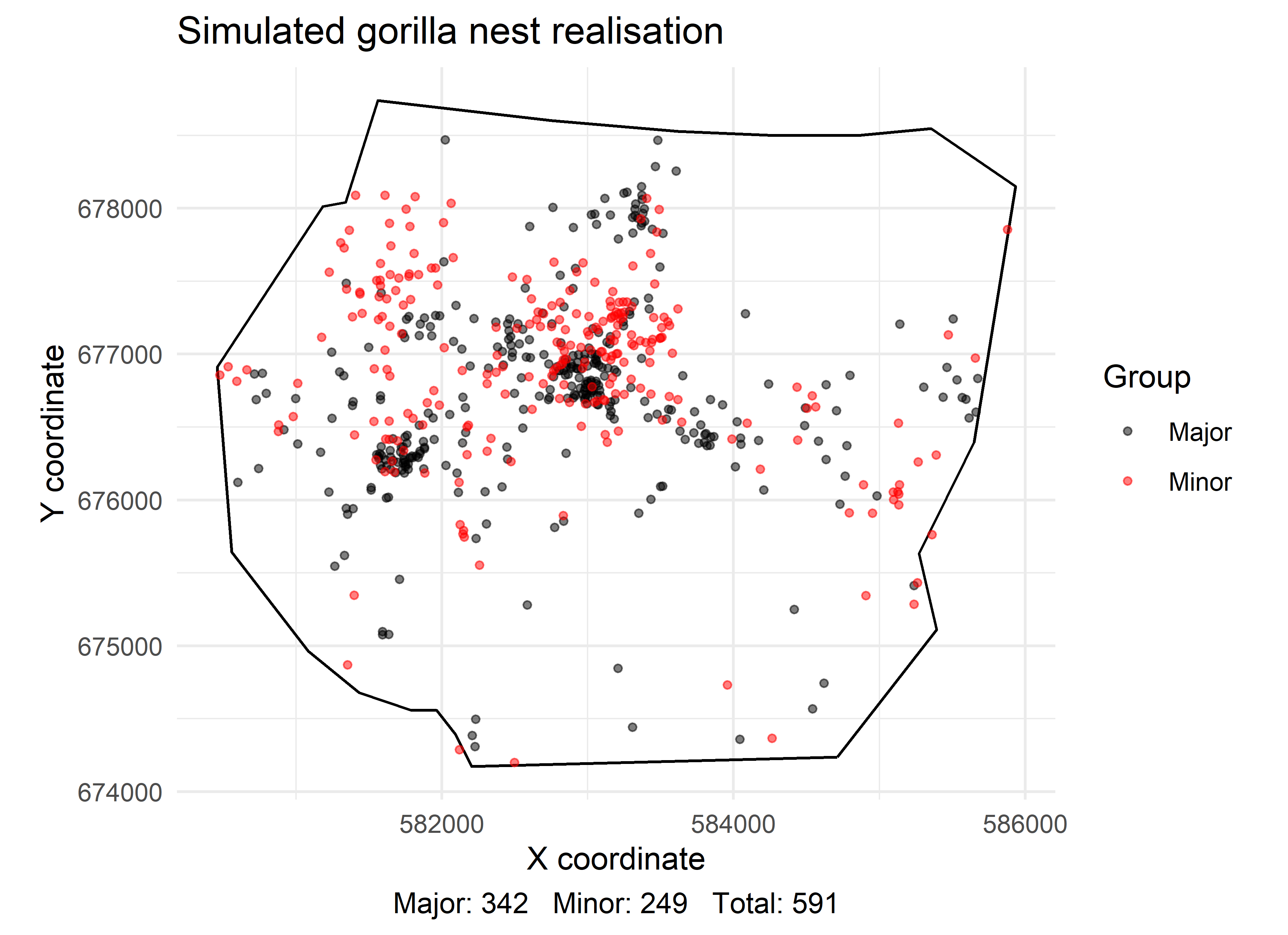}
    \includegraphics[width=0.48\linewidth]{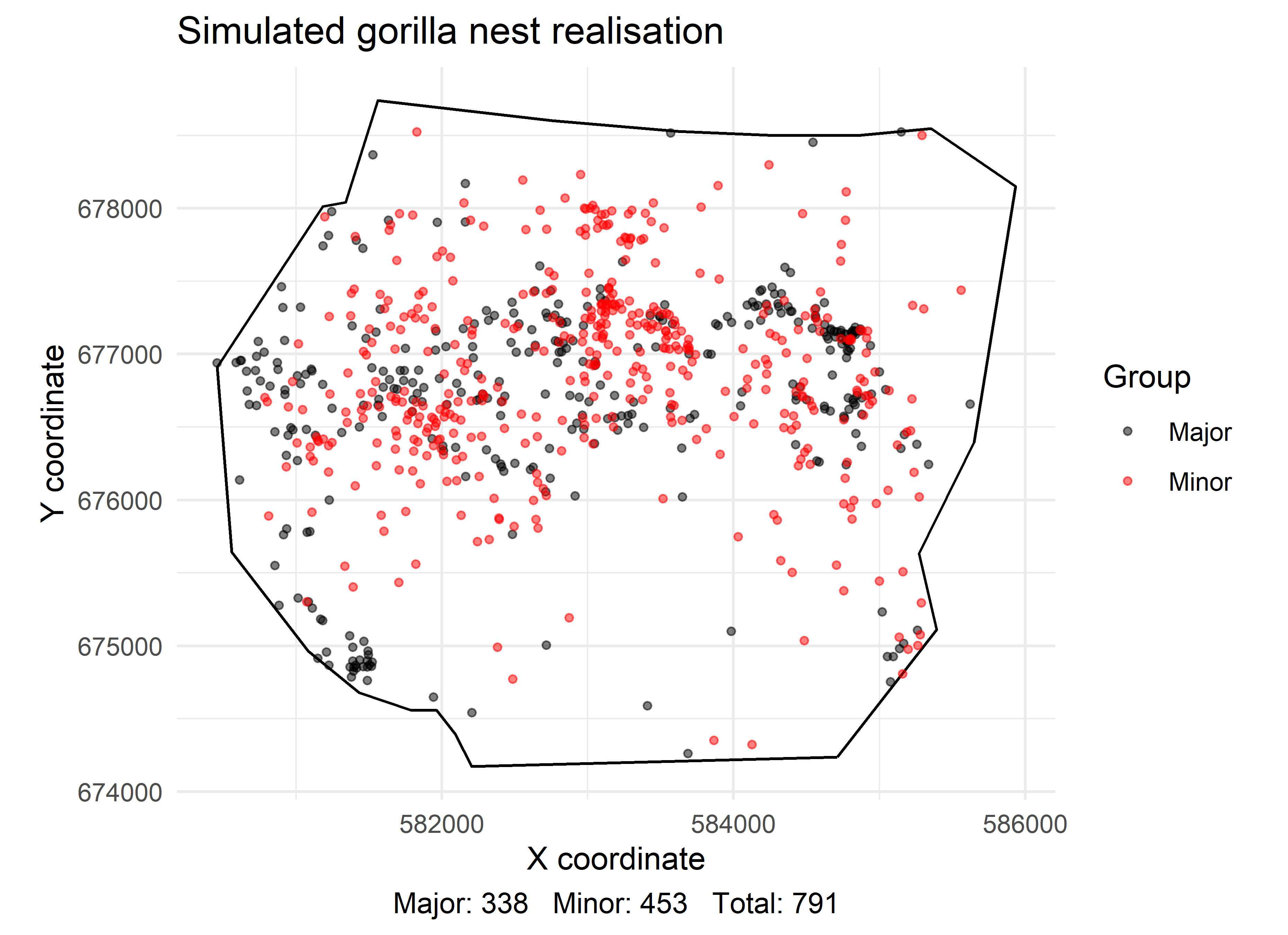}
    \vspace{0.2cm}
    \includegraphics[width=0.48\linewidth]{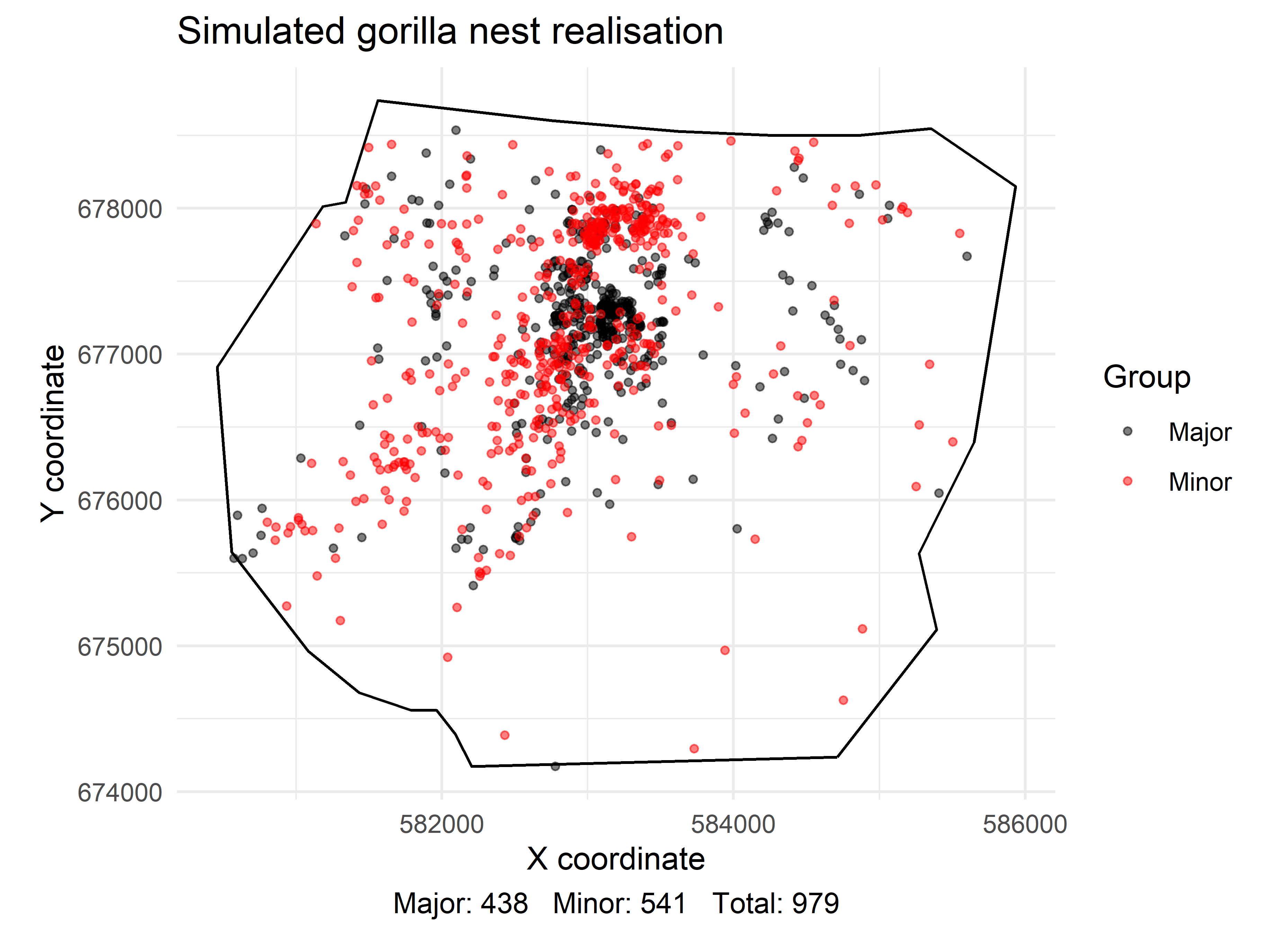}
    \includegraphics[width=0.48\linewidth]{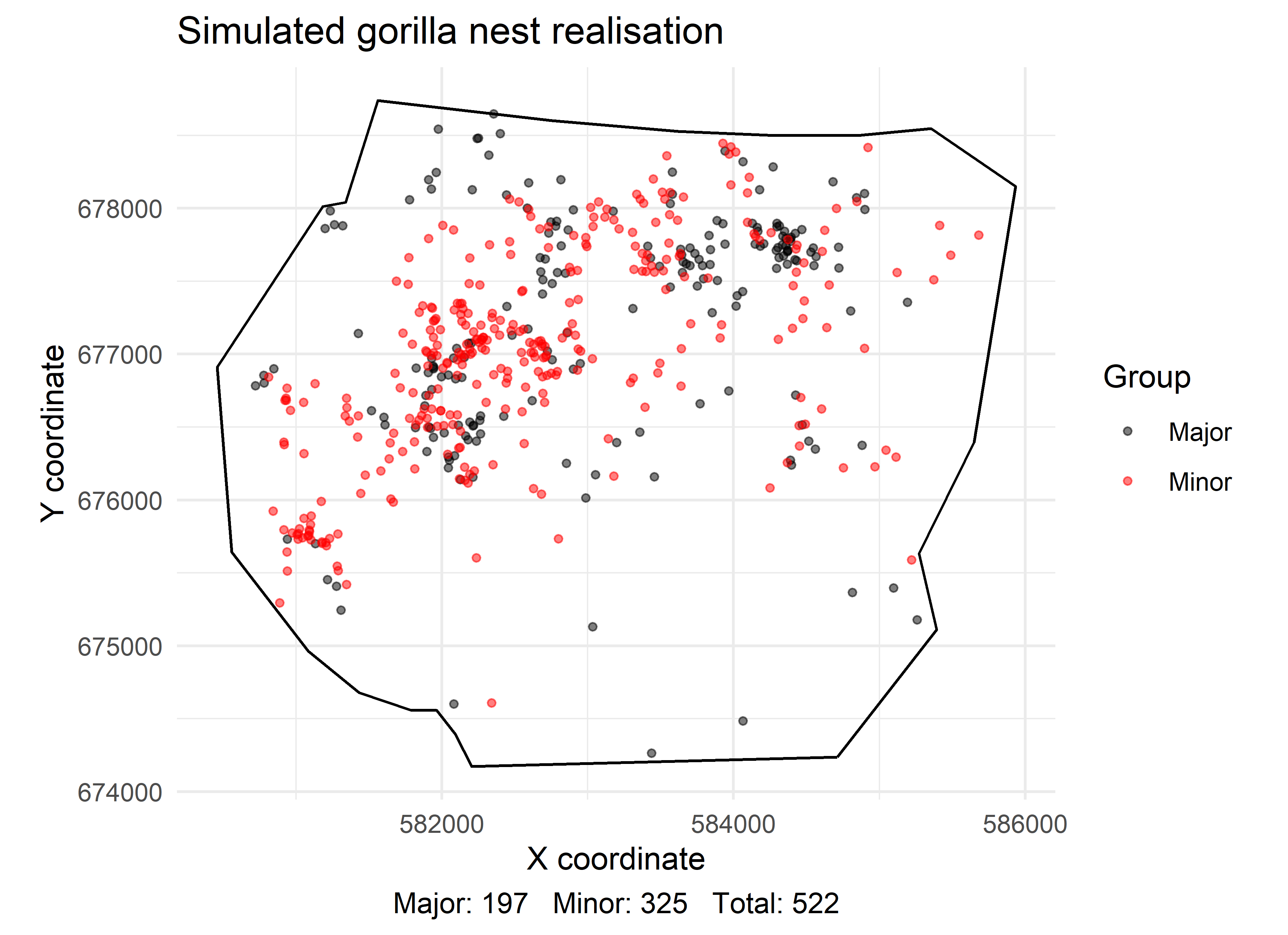}
    \caption{Examples of realisations of gorilla locations using the fitted bivariate model.}
    \label{fig:gorilla_sim_realisations}
\end{figure}

We provided examples of simulated realisations using the fitted bivariate model with the estimated parameters shown in Table~\ref{tab:mlgcp_app_estparams} over the original observation window of the gorilla data. Compared to Figure~\ref{fig:gordata}, it was clear from the four examples shown in Figure~\ref{fig:gorilla_sim_realisations} that the fitted model captured the positive correlation between the major and minor types of gorilla locations. The global locations of both types were also well captured by the fitted model. This suggested that the fitted bivariate LGCP model was adequate for the observed gorilla data.

\section{Simulation Results for Homogeneous Bivariate LGCPs with a Larger Simulation Window}
\label{sec: sim1_larger_W}
For the proposed DSBI method, increasing the observation window from $W_1=[0,1]^2$ to $W_2=[0,1.5]^2$ led to very similar performance. As shown in Figure~\ref{fig:hom_boxplot} and Table~\ref{tab:hom_comparison}, DSBI produced small biases and low RMSEs for both the variance and scale parameters under the larger window. This suggested that the proposed method was already accurate and that the homogeneous case was relatively easy to estimate. Therefore, increasing the window did not lead to a substantial additional improvement. For the MC method, the results improved when the observation window increased, which was consistent with the behaviour reported in the literature. In particular, the boxplots showed that the estimates under $W_2$ were generally closer to the true values than those under $W_1$, although the MC method still had larger variability and higher errors than the proposed method. The INLA estimates of the $\sigma$ parameters were stable and accurate for both windows. However, increasing the window did not noticeably improve the performance of INLA for the scale parameters in our setting. The INLA estimates of $\xi_Y$, $\xi_{U_1}$ and $\xi_{U_2}$ remained biased upward, with relatively large RMSE and MAPE values. This indicated that the identifiability issue for the scale parameters was not fully resolved simply by increasing the observation window.

\begin{table}[H]
\centering
\begin{tabular}{cc|ccc|ccc|ccc}
\hline
 & & \multicolumn{3}{c|}{DSBI} & \multicolumn{3}{c|}{MC} & \multicolumn{3}{c}{INLA}\\
& True   & Bias & RMSE & MAPE & Bias & RMSE& MAPE & Bias & RMSE & MAPE\\
\hline
$\sigma_Y$ & 1.5 & -0.014 &0.143 & 7.459& -0.186 &0.339 &18.899& 0.038& 0.212 &11.539\\
$\sigma_{U_1}$& 1 & 0.067 &0.200 &15.862& -0.128 &0.290 &23.802& -0.030 &0.216 &17.530\\
$\sigma_{U_2}$& 1 & 0.061 &0.199 &15.904& -0.121 &0.280 &22.674& -0.018 &0.217 &17.603\\
\hline
$\xi_Y$ & 0.2& -0.010 &0.031 &11.042& -0.011 &0.085 &37.631& 0.073 &0.076 &36.520\\
$\xi_{U_1}$& 0.15 & 0.009 &0.035& 19.427& -0.037 &0.090 &52.394& 0.096 &0.103 &64.737\\
$\xi_{U_2}$& 0.15 & 0.013 &0.036& 19.652& -0.034& 0.089& 51.252& 0.098 &0.104 &65.594\\
\hline
\end{tabular}
\caption{Comparison of parameter estimation performance for DSBI, MC and INLA under the larger observation window $W_2=[0,1.5]^2$.}
\label{tab:hom_comparison_1p5}
\end{table}

\begin{figure}[H]
    \centering
    \includegraphics[width=0.8\linewidth]{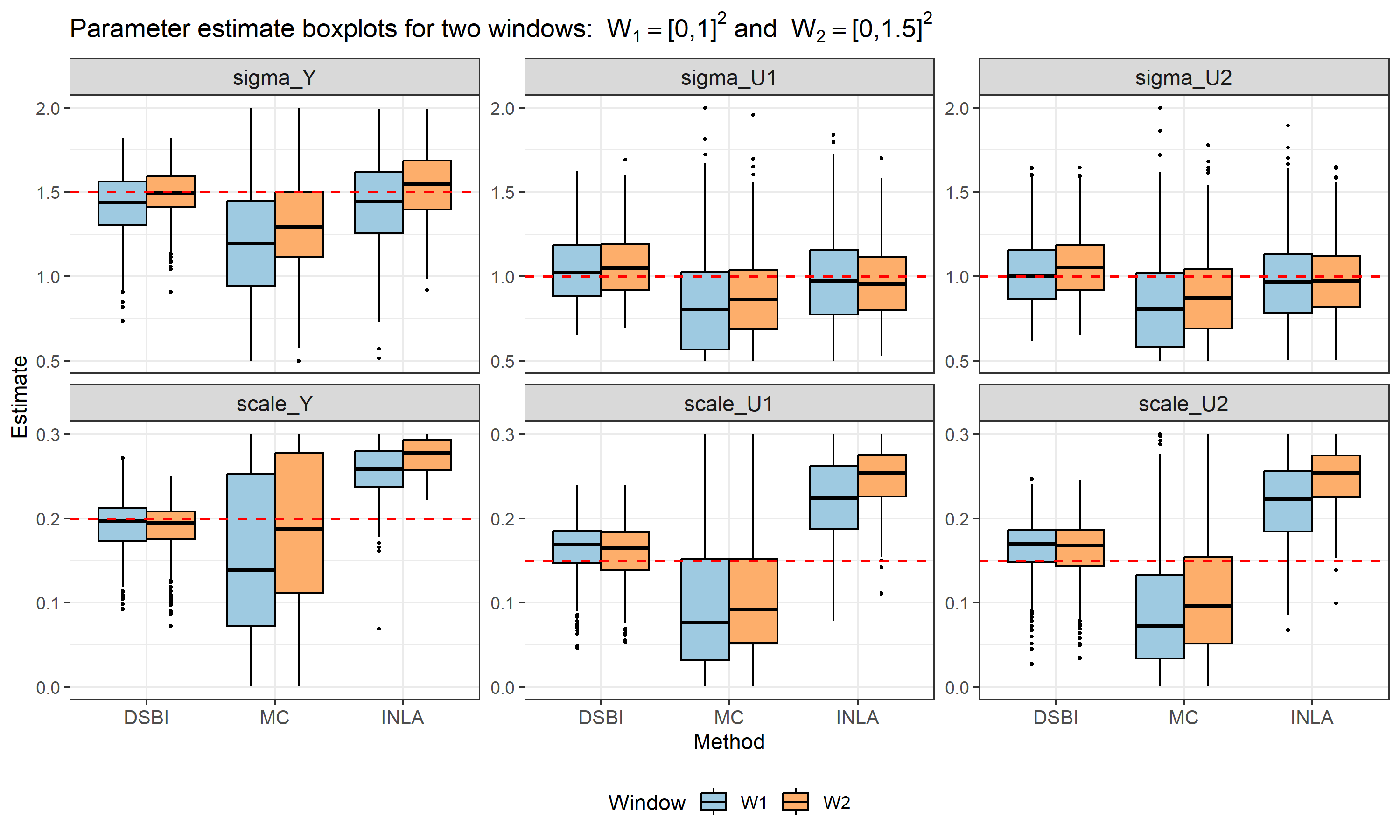}
    \caption{Boxplots of parameter estimates obtained by DSBI, MC and INLA under the two observation windows $W_1=[0,1]^2$ and $W_2=[0,1.5]^2$.}
    \label{fig:hom_boxplot_1p5}
\end{figure}

\end{document}